\documentclass[iop,numberedappendix]{emulateapj}
\usepackage{natbib}
\usepackage{graphicx}
\usepackage{rotating}
\usepackage{verbatim}
\usepackage{amssymb}

\slugcomment{Revised version, July 4th 2013 - Accepted for publication in ApJ}

\shorttitle{X-ray Nuclear Activity in S$^{4}$G Barred Galaxies}
\shortauthors{CISTERNAS ET AL.}

\begin{document}

\title{X-ray Nuclear activity in S$^{4}$G barred galaxies:\\
no link between bar strength and co-occurrent supermassive black hole fueling}
%\mbox{\hspace{-0.4cm}no relation between bar strength and co-occurrent supermassive black hole fueling}}
%bar strength has no impact on co-occurrent supermassive black hole fueling}
%bar strength has no impact on the ongoing supermassive black hole feeding}
%no impact of bar strength on the degree of ongoing supermassive black hole feeding}


\author{Mauricio Cisternas$^{1,2}$, %iac
Dimitri A. Gadotti$^{3}$, %eso
Johan H. Knapen$^{1,2}$, %iac
Taehyun Kim$^{3,4,5,6}$, %eso,nrao,seoul,carnegie
Sim\'on D\'iaz-Garc\'ia$^{7}$,\\ %oulu
Eija Laurikainen$^{7,8}$, %oulu, finca
Heikki Salo$^{7}$, %oulu
Omaira Gonz\'alez-Mart\'{\i}n$^{1,2}$, %iac
Luis C. Ho$^{6}$, %carnegie
Bruce G. Elmegreen$^{9}$,\\ %IBM
Dennis Zaritsky$^{10}$, %arizona
Kartik Sheth$^{4}$, %nrao
E. Athanassoula$^{11}$, %lam
Albert Bosma$^{11}$, %lam
S\'ebastien Comer\'on$^{7}$,\\ %oulu
Santiago Erroz-Ferrer$^{1,2}$, %iac
Armando Gil de Paz$^{12}$, %madrid
Joannah L. Hinz$^{10}$, %arizona
Benne W. Holwerda$^{13}$, %esa
Jarkko Laine$^{7}$, %oulu
Sharon Meidt$^{14}$, %mpia
Kar\'{\i}n Men\'endez-Delmestre$^{15}$, %rio
Trisha Mizusawa$^{4,16}$, %nroa,fit
Juan-Carlos Mu\~{n}oz-Mateos$^{4}$,\\ %nrao
Michael W. Regan$^{17}$, %stsci
and Mark Seibert$^{6}$ %carnegie
}
\email{mauricio@iac.es}

%% affiliation information with \altaffiltext, with one command per each affiliation.
\affil{$^{1}$ Instituto de Astrof\'{\i}sica de Canarias, E-38205 La Laguna, Tenerife, Spain}
\affil{$^{2}$ Departamento de Astrof\'{\i}sica, Universidad de La Laguna, E-38205 La Laguna, Tenerife, Spain}
\affil{$^{3}$ European Southern Observatory, Casilla 19001, Santiago 19, Chile}
\affil{$^{4}$ National Radio Astronomy Observatory, 520 Edgemont Road, Charlottesville, VA 22903, USA}
\affil{$^{5}$ Astronomy Program, Department of Physics and Astronomy, Seoul National University, Seoul 151-742, Korea}
\affil{$^{6}$ The Observatories of the Carnegie Institution for Science, 813 Santa Barbara Street, Pasadena, CA 91101, USA}
\affil{$^{7}$ Division of Astronomy, Department of Physical Sciences, University of Oulu, Oulu FIN-90014, Finland}
\affil{$^{8}$ Finnish Centre of Astronomy with ESO (FINCA), University of Turku, V\"ais\"antie 20, F1-21500, Piikki\"o, Finland}
\affil{$^{9}$ IBM T. J. Watson Research Center, 1101 Kitchawan Road, Yorktown Heights, NY 10598, USA}
\affil{$^{10}$ Steward Observatory, University of Arizona, 933 North Cherry Avenue, Tucson, AZ 85721, USA}
\affil{$^{11}$ Aix Marseille UniversitŽ, CNRS, LAM (Laboratoire d'Astrophysique de Marseille) UMR 7326, 13388, Marseille, France}
\affil{$^{12}$ Departamento de Astrof\'{\i}sica, Universidad Complutense de Madrid, E-28040 Madrid, Spain}
\affil{$^{13}$ European Space Agency, ESTEC, Keplerlaan 1, 2200 AG, Noordwijk, the Netherlands}
\affil{$^{14}$ Max-Planck-Institut f\"{u}r Astronomie, K\"{o}nigstuhl 17, D-69117 Heidelberg, Germany}
\affil{$^{15}$ Observatorio do Valongo, Universidade Federal do Rio de Janeiro, Ladeira Pedro Ant\^onio, 43, CEP 20080-090 Rio de Janeiro, Brazil}
\affil{$^{16}$ Florida Institute of Technology, Melbourne, FL 32901, USA}
\affil{$^{17}$ Space Telescope Science Institute, 3700 San Martin Drive, Baltimore, MD 21218, USA}

\begin{abstract}
Stellar bars can lead to gas inflow toward the center of a galaxy and stimulate nuclear star formation.
However, there is no compelling evidence on whether they also feed a central supermassive black hole:
by measuring the fractions of barred active and inactive galaxies, previous studies have yielded conflicting results.
In this paper, we aim to understand the lack of observational evidence for bar-driven active galactic nucleus (AGN) activity by
studying a sample of 41 nearby ($d < 35$ Mpc) barred galaxies from the {\em Spitzer} Survey for Stellar Structure in Galaxies.
We use {\em Chandra} observations to measure nuclear 2--10 keV X-ray luminosities and estimate Eddington ratios,
together with {\em Spitzer} 3.6 $\mu$m imaging to quantify the strength of the stellar bar in two independent ways:
(1) from its structure, as traced by its ellipticity and boxiness,
and (2) from its gravitational torque $Q_b$, taken as the maximum ratio of the tangential force to the mean background radial force.
In this way, rather than discretizing the presence of both stellar bars and nuclear activity, we are able to account for the continuum of bar strengths and degrees of AGN activity.
We find nuclear X-ray sources in 31 out of 41 galaxies with median X-ray luminosity and Eddington ratio of $L_{\mathrm{X}}$ = $4.3 \times 10^{38}$ erg s$^{-1}$ and $L_{\mathrm{bol}}$/$L_{\mathrm{Edd}}$ = $6.9 \times 10^{-6}$ respectively, consistent with low-luminosity AGN activity.
Including upper limits for those galaxies without nuclear detections, we find no significant correlation between any of the bar strength indicators and the degree of nuclear activity,
irrespective of galaxy luminosity, stellar mass, Hubble type, or bulge size.
Strong bars do not favor brighter or more efficient nuclear activity, implying that at least for the low-luminosity regime, supermassive black hole fueling is not closely connected to large scale features. 
\end{abstract}



\section{Introduction}

Supermassive black holes (BHs), expected to reside in the centers of most massive galaxies \citep{mbh_l1,richstone98}, experienced the bulk of their growth around 10 billion years ago in short periods of vigorous mass accretion \citep{lynden-bell69, soltan82, yu&tremaine02, ueda03, marconi04, shankar04}.
During these phases, BHs can be observed as quasars, the extremely bright-end of the active galactic nucleus (AGN) family.
While compared to earlier times, our present-day universe can be considered quiescent in terms of BH activity, there happens to be a very significant fraction of nearby galaxies showing some level of AGN activity:
the Palomar spectroscopic survey of local galaxies revealed that $\sim$40\% of them display nuclear activity likely due to BH fueling \citep{ho97a}, yet they represent the faint-end of the AGN luminosity function and feature very modest accretion rates \citep{ho09b}.

The mechanisms through which these BHs are fed are still a matter of investigation \citep[for reviews, see][]{wada04,martini04}.
The basic requirement is that a fraction of the galaxy's interstellar medium, distributed over kiloparsec scales, has to be deprived of its angular momentum in such a way that is able to reach the innermost regions of the galaxy, close to the BH.
Secular processes, i.e., those that take longer than a dynamical timescale to be relevant \citep[for a review, see][]{kormendy&kennicutt04}, are expected to be the dominant mechanisms feeding low-luminosity AGNs (LLAGNs) and even moderate luminosity ones out to $z\sim 1$ \citep[e.g.,][]{gabor09, georgakakis09, cisternas11a, cisternas11b} and even $z\sim 2$ \citep[e.g.,][]{jahnke09, bennert11,kocevski12}.
In this regard, non-axisymmetric structures such as stellar bars can lead to internal instabilities and gas inflows, the necessary elements to bring gas to the center and, perhaps, fuel the BH.

Bars can play a major role in the overall evolution of a galaxy by driving its gaseous interstellar medium toward its inner regions.
Through their non-axisymmetric potential, large-scale stellar bars exert torques that accumulate gas and dust at the leading end of the bar where they get shocked, lose angular momentum, and fall inward the central regions of the galaxy \citep{athanassoula92b, knapen95, regan99b, maciejewski02, sheth02, kim12b}.
A diversity of observational studies support this picture:
gas kinematics have revealed streaming motions inward along the bar \citep[e.g.,][]{regan97, mundell99, erroz12};
against their unbarred counterparts, barred galaxies show higher central concentrations of molecular gas \citep{sakamoto99, sheth05},
as well as enhanced nuclear star formation rates \citep[e.g.,][]{devereux87, hummel90, martin95, ho97b, sheth00, ellison11, wang12},
and a higher rate of bulges with young stellar populations \citep{coelho11}.
The possibility that these bar-induced inflows could reach the central few parsecs and fuel a BH led large-scale bars to be proposed early-on as a plausible mechanism to trigger AGN activity \citep{simkin80, shlosman89}.

While the spatial scales involved in transporting gas from a large-scale bar to a central BH differ by a few orders of magnitude, {\em Hubble Space Telescope (HST)} observations of the central regions of barred galaxies revealed nuclear dust spiral structure connecting the kiloparsec-scale bar all the way down to the central tens of parsecs, at the resolution limit of these observations \citep{martini03a}.
These nuclear dust structures tend to be found in a minority of galactic centers, and with comparable frequencies on both active and inactive galaxies.
This suggests that (1) the AGN lifetime is less than the inflow time of these spiral structures and (2) no unique fueling mechanism can be traced at these intermediate spatial scales \citep{martini03b}.
On the other hand, most current dynamical models agree on long-lived stellar bars \citep[see, e.g.,][]{athanassoula13}, and therefore if the gas being currently consumed by an active BH was initially driven by a large-scale stellar bar, one would expect {\em some} correlation between bars and galaxies with {\em ongoing} nuclear activity.

A number of studies have searched for the appealing ``bar-AGN connection'', mainly by looking at samples of active and inactive galaxies and measuring their bar fractions, or conversely, by studying the AGN fraction among samples of barred and unbarred galaxies.
Results have been mixed:
while some studies have found tentative evidence in favor of an observable link between barred galaxies and AGNs \citep{arsenault89, knapen00, laine02, maia03, laurikainen04a, coelho11,oh12},
others have not found a causal connection between the presence of bars and AGN activity \citep{moles95, mcleod95, mulchaey97, ho97b, hunt&malkan99, lee12},
and others have even found hints for an anti correlation between the presence of a bar and nuclear activity \citep{shlosman00, zhang09}.

In general, the aforementioned studies tend to discretize either (or both) bars or AGNs.
Stellar bars can have a wide range of properties which will define their strength:
a strong bar will induce a different level of inflow than a weak bar.
AGNs, on the other hand, have a continuous distribution in luminosity and BH accretion rate that spans a few orders of magnitude implying that there are very different levels of nuclear activity.
In this paper we explore the possibility of a hitherto overlooked link between bar strength and degree of AGN activity.
We select a sample of barred galaxies from the {\em Spitzer} Survey of Stellar Structure in Galaxies \citep[S$^{4}$G,][]{s4g} and take advantage of the 3.6 $\mu$m imaging, a reliable tracer of the old stellar population which makes up the bar, to characterize its strength through its structural properties and relative torque.

To characterize the level of BH activity, we opt to use {\em Chandra} X-ray observations, which offer a number of advantages with respect to optical diagnostics when attempting to uniformly study the low-luminosity regime.
While optical emission lines such as H$\alpha$ can suffer from contamination from extranuclear sources not related to the central engine, X-ray emission originates much closer to where the accretion is taking place, and given {\em Chandra}'s high resolution, one can identify and isolate the X-ray nuclear source from other sources of emission.
X-ray observations have proven to be highly efficient in revealing previously undetected AGNs \citep[e.g.,][]{martini02, tzanavaris07, pellegrini07, gallo08, ghosh08, grier11}, most notably in galaxies lacking classical bulges and conventionally thought to be unlikely BH (and hence AGN) hosts \citep[e.g.,][]{desroches&ho09, araya12}.

Starting from a sample of S$^{4}$G barred galaxies, as described in Section 2, we gather all the available archival {\em Chandra} data.
In Section 3 we present the X-ray data analysis and asses the level of nuclear activity, and in Section 4 we describe how the strength of the stellar bars was quantified.
We report our results in Section 5 and discuss their implications within the context of previous findings from the literature in Section 6.

\begin{deluxetable*}{lr@{.}lr@{.}lcccr@{.}lccr@{}lr@{.}lcr@{}lc}
\tabletypesize{\scriptsize}
\tablecaption{Sample details and nuclear properties\label{tab1}}
\tablehead{Galaxy & \multicolumn{2}{c}{$d$} & \multicolumn{2}{c}{$T$-type} & Nuclear & Spec. & Obs.ID & \multicolumn{2}{c}{$t_{exp}$} & X-ray & Counts & \multicolumn{2}{c}{log $L_{\mathrm{X}}$} \\
 & \multicolumn{2}{c}{(Mpc)} & \multicolumn{2}{c}{} & Bar/Ring& Class & & \multicolumn{2}{c}{(ks)} & Class &  & \multicolumn{2}{c}{(erg s$^{-1}$)} \\
 (1) & \multicolumn{2}{c}{(2)} & \multicolumn{2}{c}{(3)} &(4) &(5) & (6) & \multicolumn{2}{c}{(7)} & (8) & (9) & \multicolumn{2}{c}{(10)} 
 }
\startdata
NGC 255  &   20&0 &   4&0 &... &  ... &   7844 &   4&6 & IV & $<$2 & $<$&  36.4 \\
NGC 685  &   15&1 &   5&4 & ... &... &   7857 &   4&6 & IV & $<$3 & $<$&  37.8 \\
NGC 1036  &   11&1 &   0&0 &... & ... &   7119 &   3&0 & IV & $<$2 & $<$&  36.5 \\
NGC 1073  &   15&1 &   5&3 &... & ... &   4686 &   5&6 & I & 27 & &  38.3 \\
NGC 1097  &   20&0 &   3&2 & nb,nr & L &   2339 &   5&6 & II & 1828 & &  40.9 \\
NGC 1232  &   18&6 &   5&0 &... & ... &  10798 &  52&9 & I & 81 & &  38.6 \\
NGC 1291  &    8&6 &   0&1 & nb & L &  11272 &  69&0 & II & 812 & &  39.2 \\
NGC 1300  &   18&0 &   4&0 & nr & ... &  11775 &  29&7 & II & 125 & &  38.6 \\
NGC 1302  &   20&0 &   0&1 &... & ... &   7847 &   4&9 & I & 13 & &  37.3 \\
NGC 1341  &   16&8 &   1&2 &... & ... &   7846 &   4&9 & IV & $<$2 & $<$&  35.9 \\
NGC 1367  &   23&2 &   1&1 &... & ... &   7277 &  14&8 & I & 455 & &  40.7 \\
NGC 1493  &   11&3 &   6&0 &... & ... &   7145 &  10&0 & I & 51 & &  38.5 \\
NGC 1637  &   10&6 &   5&0 &... & ... &    766 &  38&5 & I & 179 & &  38.2 \\
NGC 1640  &   19&1 &   3&0 &... & ... &   7891 &   5&0 & I & 23 & &  38.6 \\
NGC 1672  &   14&5 &   3&2 &nr & S &   5932 &  39&5 & II & 91 & &  38.4 \\
NGC 2787  &   10&2 &  -1&0 &nr & L &   4689 &  30&7 & I & 500 & &  40.0 \\
NGC 3344  &    6&0 &   4&0 &... & ... &   7087 &   1&7 & I & 12 & &  38.0 \\
NGC 3351  &   10&1 &   3&0 & nr& ... &   5931 &  39&5 & III & $<$32 & $<$&  37.3 \\
NGC 3368  &   10&8 &   2&2 & nb,nr & ... &    391 &   2&0 & II & 8 & &  36.3 \\
NGC 3627  &   10&0 &   3&0 &... &  L &   9548 &  49&5 & II & 48 & &  37.8 \\
NGC 4136  &    9&6 &   5&1 &... & ... &   2921 &  19&7 & I & 15 & &  37.4 \\
NGC 4245  &    9&6 &   0&1 & nr & ... &   7107 &   2&2 & IV & $<$6 & $<$&  38.2 \\
NGC 4303  &   16&4 &   4&0 & nb,nr& S2 &   2149 &  28&0 & II & 154 & &  38.8 \\
NGC 4314  &    9&6 &   1&0 & nr & ... &   2062 &  16&1 & II & 26 & &  37.8 \\
NGC 4394  &   16&7 &   3&0 &... & ... &   7864 &   5&0 & III & $<$8 & $<$&  36.2 \\
NGC 4450  &   16&5 &   2&4 &... & L &   3997 &   3&4 & I & 479 & &  40.2 \\
NGC 4548  &   16&2 &   3&0 &... & L &   1620 &   2&7 & I & 27 & &  38.6 \\
NGC 4579  &   19&5 &   2&9 & nr& S2 &    807 &  33&9 & II & 26437 & &  41.4 \\
NGC 4596  &   16&7 &   0&1 & ...& ... &  11785 &  31&0 & I & 45 & &  38.3 \\
NGC 4639  &   22&3 &   3&5 &... & S1 &    408 &   1&3 & I & 417 & &  42.0 \\
NGC 4713  &   16&3 &   6&8 &... & ... &   4019 &   4&9 & I & 9 & &  38.3 \\
NGC 4725  &   13&6 &   2&2 & nb& S2 &   2976 &  24&6 & I & 397 & &  41.1 \\
NGC 5350  &   31&2 &   3&5 &... & ... &   5903 &   4&5 & II & 13 & &  38.8 \\
NGC 5371  &   29&4 &   4&0 &... & S &  13006 &   5&4 & I & 19 & &  39.2 \\
NGC 5584  &   26&7 &   6&0 &... & ... &  11229 &   7&0 & IV & $<$3 & $<$&  36.0 \\
NGC 5728  &   30&5 &   1&2 & nb,nr& S1.9 &   4077 &  18&7 & II & 503 & &  40.2 \\
NGC 5964  &   26&5 &   6&9 &... & ... &  12982 &   9&8 & IV & $<$2 & $<$&  37.8 \\
NGC 7479  &   33&8 &   4&3 &... & S1.9 &  11230 &  24&7 & I & 105 & &  39.0 \\
NGC 7552  &   17&1 &   2&4 & nr & H2 &   7848 &   5&0 & II & 131 & &  39.1 \\
NGC 7743  &   21&4 &   0&1 &... & S2 &   6790 &  13&8 & I & 88 & &  38.4 \\
PGC 3853  &   12&6 &   7&0 &... & ... &  12981 &   9&8 & IV & $<$6 & $<$&  37.7
\enddata
\tablecomments{
Column (1): galaxy name;
Column (2): redshift independent distances from NED;
Column (3): morphological $T$-type from HyperLeda \citep{hyperleda};
Column (4): nuclear bars (nb) and nuclear rings (nr) are indicated if present, based on \citet{buta10}, \citet{laine02}, \cite{erwin04}, and \citet{comeron10};
Column (5): nuclear spectroscopic classification from \citet{ho97a}, \citet{veron06}, and \citet{smith07}. where L=LINER, S=Seyfert, and its attached number indicating its particular type, and H2=H {\sc ii} nucleus;
Column (6): {\em Chandra} observation ID;
Column (7): net exposure time of processed observation;
Column (8): X-ray nuclear classification as illustrated in Figure \ref{figx};
Column (9): effective background-subtracted broadband counts or upper limit when no nuclear point source was detected;
Column (10): Intrinsic X-ray luminosity in the 2--10 keV band or upper limit.
%Methods used for direct black hole mass measurements: NGC 1300, gas kinematics \citep{atkinson05}; NGC 2787 and NGC 4596, gas kinematics \citep{sarzi01}; and NGC 3368, stellar kinematics \citep{nowak10}.
}
\end{deluxetable*}

\section{Sample and Data}

In this paper we analyze a sample of barred galaxies drawn from the S$^{4}$G data set with the goal of studying whether bar strength and X-ray nuclear activity are connected.
Below we briefly describe the S$^{4}$G survey, as well as the parent sample of barred galaxies for which we searched for archival {\em Chandra} X-ray data.

\subsection{S$^{4}$G Barred Galaxy Sample}

S$^{4}$G is a post-cryogenic Cycle 9 Science Exploration Program aiming to provide near-infrared (NIR) imaging of over 2300 nearby ($d < 40$ Mpc) galaxies at 3.6 and 4.5 $\mu$m with the Infrared Array Camera \citep[IRAC;][]{irac} onboard the {\em Spitzer Space Telescope}.
Galaxy images are uniformly processed with the S$^{4}$G reduction pipeline \citep[for details, see][]{s4g}, with the final mosaics having a 0\farcs75/pixel scale and a resolution of 1\farcs7.

In this paper, we use the parent sample of barred galaxies selected by T.~Kim et al. (in preparation), in which the structure of stellar bars is explored in detail.
At the time of the sample selection (November 2011), over 50 percent of the S$^4$G sample had already been processed by the basic pipelines, providing science-ready images.
Barred galaxies were identified as such through a visual inspection of the NIR images by members of the S$^4$G team.
From these, a total of 144 barred galaxies were selected based on the following criteria:
sample selection focussed firstly on avoiding doubtful cases,
galaxies which were highly inclined ($b/a>0.5$), significantly disturbed by an
ongoing close interaction or merger, overly faint or irregular, or
simply unsuitable for image fitting (e.g. because of a bright foreground star in a critical position).
Secondly, the selection was done in a way in which a good coverage of all disk Hubble types is obtained.
While this means that the sample is not complete, these selection procedures assure that the sample is
(1) representative of the local population of barred galaxies,
and (2) suitable for structural analysis via image decomposition,
meaning that the structural parameters can be accurately derived.

%%%%%%%%%%%%%%%%%%%%%%%%
\begin{figure}[t]
\centering
\resizebox{\hsize}{!}{\includegraphics{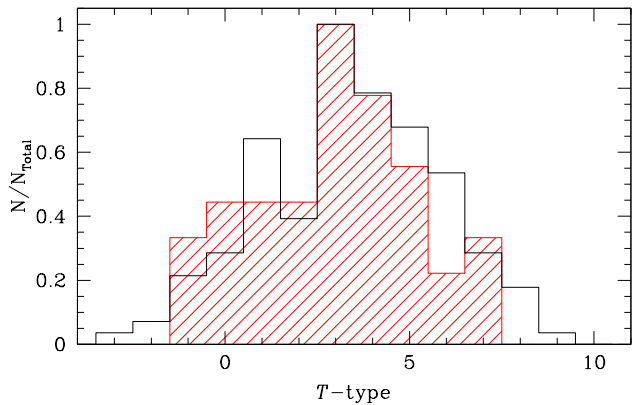}}
\caption{Normalized distributions of morphological $T$-type for the parent sample of barred galaxies (unfilled histogram) as well as for the {\em Chandra} subsample (shaded histogram).\label{figtt}}
\end{figure}
%%%%%%%%%%%%%%%%%%%%%%%%

\subsection{Chandra X-ray Data}

To identify possible X-ray emission from AGNs we look for archival observations carried out with the Advanced CCD Imaging Spectrometer \citep[ACIS;][]{acis} onboard {\em Chandra}.
Compared to other X-ray observing facilities, ACIS offers excellent angular resolution, featuring a point-spread function (PSF) with a full width at half maximum of 1$^{\prime\prime}$.
This allows us to search for point like emission coincident with the NIR center, and at the same time avoid confusion with other X-ray emitting sources, such as surrounding diffuse hot gas and unresolved X-ray binaries.

Forty-one out of 144 galaxies from our parent sample have publicly available ACIS observations from the {\em Chandra} Data Archive\footnote{http://cxc.harvard.edu} (as of October 2012).
When more than one observation was available for a given galaxy, the one with the longest exposure time was selected.
ACIS consists of 10 CCDs arranged in two configurations: a 2$\times$2 array (ACIS-I) and a 1$\times$6 array (ACIS-S), with the former designed for imaging, and the latter used for both imaging and grating spectroscopy.
Out of the 41 ACIS observations, 39 were carried out with the S-array and the remaining two (NGC 1232 and NGC 5350) with the I-array.
The data analysis is performed on data collected by the on-axis chips, meaning both S2 and S3 chips for the S-array observations and all of the four I chips when the I-array was used.

Given that these are archival observations from individual programs and not part of a uniform X-ray survey, we could in principle be biased towards X-ray luminous active galaxies.
Nevertheless, not all of the {\em Chandra} observations analyzed here were designed to study nuclear activity in nearby galaxies.
Many of these galaxies were observed with the aim of studying supernovae or other ultra luminous X-ray sources (ULXs), and hence a wide range of luminosities is expected.
We also expect a random sampling in terms of morphological type, but to check whether a bias exists, in Figure \ref{figtt} we compare the distribution of $T$-types of the parent sample of barred galaxies to the {\em Chandra} sample, finding that both distributions roughly agree with each other and no significant bias should be present.

%%%%%%%%%%%%%%%%%%%%%%%%
\begin{figure*}[th]
\centering
\resizebox{0.95\hsize}{!}{\includegraphics{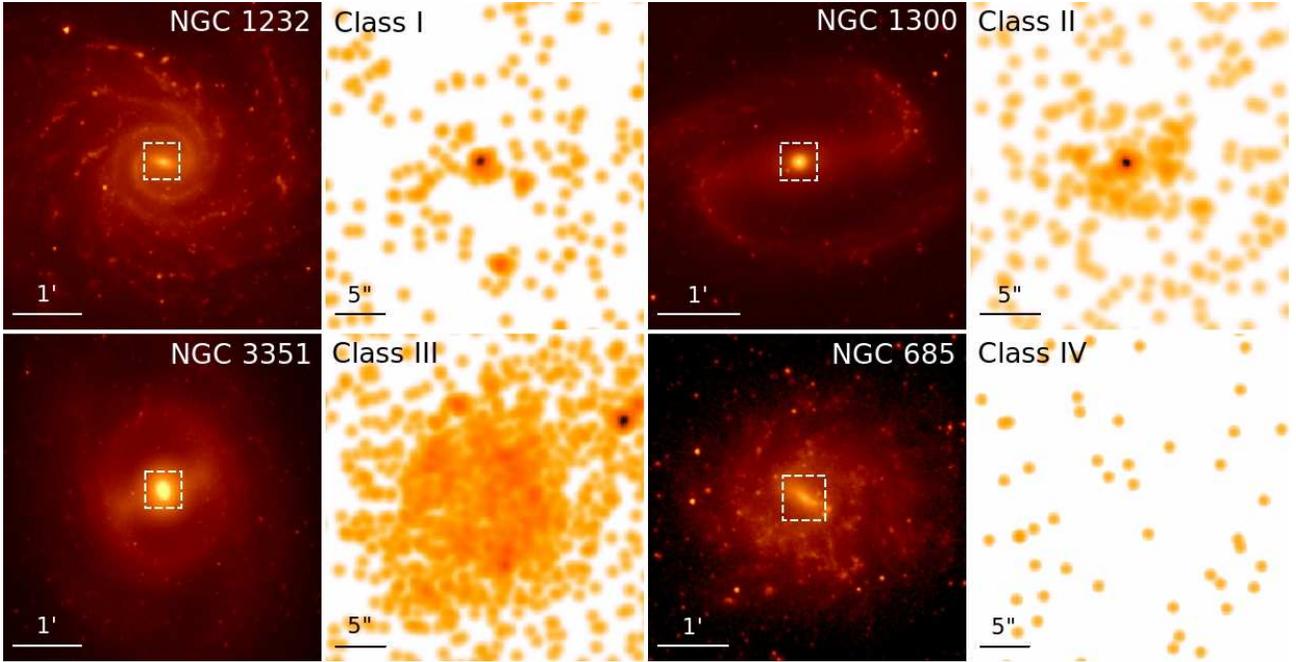}}
\caption{
Near-infrared and X-ray images of four example barred galaxies illustrating the different classes of nuclear X-ray morphologies, based on the classification by \citet{zhang09}:
(I) a dominant point source, (II) a point source embedded in diffuse emission, (III) extended diffuse emission without a distinguishable point source, and (IV) no clear emission above the background level.
For each of these four cases, the left panel shows the {\em Spitzer}/IRAC 3.6 $\mu$m image of the galaxy processed through the S$^{4}$G pipeline \citep{s4g}.
The dashed box indicates the area shown on the right, which corresponds to the {\em Chandra}/ACIS 0.2--10 keV smoothed image of the central region of the galaxy.
%Examples of the four classes of X-ray nuclear morphologies (which I will briefly explain again in this caption). For each of the four classes, we show the galaxy imaged with {\em Spitzer}/IRAC in 3.6 $\mu$m (left) together with the {\em Chandra}/ACIS 0.2--10 keV smoothed image (right).
\label{figx}}
\end{figure*}
%%%%%%%%%%%%%%%%%%%%%%%%

\section{X-ray Data Analysis}

The retrieved {\em Chandra}/ACIS level 2 event files were processed and analyzed uniformly using the {\em Chandra} Interactive Analysis of Observations (CIAO) v4.3, following the reduction procedure described in \citet{ogm06}.
We reprocessed each observation to account for possible background flares that could incorrectly enhance the count rate of our sources using the task {\ttfamily lc\_clean.sl}.
To identify the extraction regions in the ACIS images, we used the positions from the Two Micron All Sky Survey \citep[2MASS;][]{2mass} for all galaxies in our sample except for NGC 5964, for which we use the position from \citet{leon03}.
%The nuclear position from the 3.6 $\mu$m IRAC image, determined by  was used as a prior to identify the extraction region in the ACIS image.

Defining the source and background regions will depend on the morphology of the X-ray emission in the nuclear region.
Following the classification from \citet{zhang09}, we group the nuclear morphologies into four distinct classes:
(I) a dominant point source,
(II) a nuclear point source embedded in extended diffuse emission,
(III) extended emission without a point source,
and (IV) no nuclear source or diffuse emission present above the background level.
In Figure \ref{figx} we give representative examples of these four classes.

For class I nuclei, we defined the source region as a 2$^{\prime\prime}$ radius circular aperture.
The background region was defined as a source-free circular annulus around the nuclear position,
or alternatively, as several circular apertures if other point-like sources were present in the vicinity.
In the case of class II nuclei, source regions were defined as 1.5--2$^{\prime\prime}$ radius apertures depending on the extension of the surrounding diffuse emission.
We carefully defined the background to correctly characterize the spatial variations of the diffuse emission in which the nuclear source is embedded.
For classes III and IV, in which there is no distinct point source, we derive upper limits on the nuclear source by considering the background-subtracted counts within a 2$^{\prime\prime}$ aperture at the photometric center.
In a few cases with class IV nuclei and low exposure times, no counts were detected with the standard aperture size and hence larger apertures of up to 5$^{\prime\prime}$ were required to compute upper limits.

\subsection{X-ray Luminosities}

For all but nine sources (see below) we estimate X-ray luminosities in the hard 2--10 keV band (hereafter referred to as $L_{\mathrm{X}}$) by using a single power law model with Galactic interstellar absorption obtained using the {\ttfamily nH} task included in {\ttfamily FTOOLS} \citep{nh1,nh2}.
We assumed a typical photon index for low luminosity AGNs of $\Gamma=1.8$ \citep{ho01}, which has been shown to derive reliable luminosities when compared against results from a dedicated spectral fitting \citep{ogm06}.
All parameters are held fixed expect for the power-law normalization, which is found by fitting the aforementioned model using {\ttfamily XSPEC} v12.7.0 \citep{xspec}.

Nine sources in our sample present more than 200 net counts and allow for a detailed characterization of their spectra.
For these objects, we perform a spectral analysis using {\ttfamily XSPEC} following the approach by \citet{ogm09}, in which an ensemble of models including both thermal and non-thermal components is used to better dissect the true nature of our sources contributing to the observed nuclear emission.
Details on the modeling and results from the spectral fitting are presented in the Appendix, and the resulting 2--10 keV X-ray luminosities for the whole sample are presented in Table \ref{tab1}.

\begin{deluxetable*}{lr@{.}lccccr@{}l}   %ccr@{.}lccr@{}lr@{.}lcr@{}lc}
\tabletypesize{\scriptsize}
\tablecaption{Black hole masses and Eddington ratios\label{tab2}}
\tablehead{
Galaxy & \multicolumn{2}{c}{$\sigma_\ast$} & Ref. & $M^{3.6}$ & $B/T$ & log $M_{\mathrm{BH}}$ & \multicolumn{2}{c}{log $L_{\mathrm{bol}}/L_{\mathrm{Edd}}$} \\
              & \multicolumn{2}{c}{(km s$^{-1}$)}  &         & (AB)            &        & ($M_{\odot}$)                   & \multicolumn{2}{c}{} \\
 (1)        & \multicolumn{2}{c}{(2)}                     & (3)   & (4)               & (5)  & (6)                                       & \multicolumn{2}{c}{(7)}
}
\startdata
NGC 255 & . & . & ... & -19.26 &  0.00 & ... & &... \\
NGC 685 & . & . & ... & -19.36 &  0.00 & ... & &... \\
NGC 1036 & . & . & ... & -17.20 &  0.00 & ... & &... \\
NGC 1073 &  24 & 7 &        1 & -19.62 &  0.00 &    4.3 & &  -2.9 \\
NGC 1097 & 196 & 0 &        2 & -22.74 &  0.24 &    8.1 & &  -4.2 \\
NGC 1232 & . & . & ... & -21.50 &  0.04 &    6.9 & &  -5.3 \\
NGC 1291 & 186 & 0 &        3 & -21.46 &  0.38 &    8.0 & &  -5.8 \\
NGC 1300 & . & . & ... & -21.06 &  0.11 &    7.8$^{a}$ & &  -6.2 \\
NGC 1302 & 158 & 0 &        3 & -21.06 &  0.39 &    7.7 & &  -7.3 \\
NGC 1341 &  80 & 4 &        4 & -18.95 &  0.00 &    6.4 & $<$&  -7.5 \\
NGC 1367 & . & . & ... & -21.47 &  0.13 &    7.4 & &  -3.6 \\
NGC 1493 &  25 & 0$^{b}$ &        5 & -18.82 &  0.00 &    4.3 & &  -2.8 \\
NGC 1637 & . & . & ... & -19.52 &  0.12 &    6.6 & &  -5.4 \\
NGC 1640 & . & . & ... & -20.03 &  0.25 &    7.1 & &  -5.4 \\
NGC 1672 & 110 & 0 &        6 & -21.45 &  0.28 &    7.0 & &  -5.6 \\
NGC 2787 & . & . & ... & -20.06 &  0.42 &    7.6$^{c}$ & &  -4.5 \\
NGC 3344 &  73 & 5 &        1 & -18.90 &  0.06 &    6.3 & &  -5.3 \\
NGC 3351 & 119 & 9 &        1 & -20.82 &  0.19 &    7.2 & $<$&  -6.8 \\
NGC 3368 & . & . & ... & -21.33 &  0.29 &    6.9$^{d}$ & &  -7.5 \\
NGC 3627 & 124 & 0 &        1 & -21.66 &  0.12 &    7.2 & &  -6.4 \\
NGC 4136 &  38 & 4 &        1 & -18.15 &  0.03 &    5.1 & &  -4.7 \\
NGC 4245 &  82 & 6 &        1 & -19.03 &  0.36 &    6.5 & $<$&  -5.2 \\
NGC 4303 &  84 & 0 &        1 & -21.53 &  0.09 &    6.5 & &  -4.7 \\
NGC 4314 & 117 & 0 &        1 & -19.86 &  0.33 &    7.1 & &  -6.2 \\
NGC 4394 & 115 & 5 &        1 & -20.44 &  0.26 &    7.1 & $<$&  -7.9 \\
NGC 4450 & 135 & 0 &        1 & -21.32 &  0.16 &    7.4 & &  -4.1 \\
NGC 4548 & 113 & 4 &        1 & -21.31 &  0.25 &    7.1 & &  -5.4 \\
NGC 4579 & 165 & 0 &        1 & -22.32 &  0.17 &    7.8 & &  -3.3 \\
NGC 4596 & . & . & ... & -21.19 &  0.28 &    7.9$^{c}$ & &  -6.6 \\
NGC 4639 &  96 & 0 &        1 & -20.40 &  0.17 &    6.8 & &  -1.7 \\
NGC 4713 &  23 & 2 &        1 & -19.24 &  0.00 &    4.2 & &  -2.8 \\
NGC 4725 & 140 & 0 &        1 & -21.73 &  0.19 &    7.5 & &  -3.3 \\
NGC 5350 & . & . & ... & -21.09 &  0.07 &    7.0 & &  -5.1 \\
NGC 5371 & 179 & 8 &        1 & -22.00 &  0.12 &    7.9 & &  -5.7 \\
NGC 5584 & . & . & ... & -20.27 &  0.00 & ... & &... \\
NGC 5728 & 209 & 0 &        3 & -21.76 &  0.28 &    8.2 & &  -5.0 \\
NGC 5964 & . & . & ... & -20.27 &  0.00 & ... & &... \\
NGC 7479 & 154 & 6 &        1 & -22.30 &  0.12 &    7.7 & &  -5.6 \\
NGC 7552 & 104 & 0 &        7 & -21.31 &  0.37 &    6.9 & &  -4.8 \\
NGC 7743 &  89 & 3 &        1 & -20.47 &  0.31 &    6.6 & &  -5.2 \\
PGC 3853 & . & . & ... & -18.93 &  0.00 & ... & &...
\enddata
\tablecomments{
Column (1): galaxy name;
Column (2): central stellar velocity dispersion;
Column (3): Reference for either $\sigma_{\ast}$ or $M_{\mathrm{BH}}$:
(1) \citet{ho09a};
(2) \citet{lewis06};
(3) \citet{mcelroy05};
(4) \citet{wegner03};
(5) \citet{walcher05};
(6) \citet{garcia-rissmann05}
(7) \citet{oliva95}.
Column (4): absolute magnitude from \citet{munoz-mateos13};
Column (5): bulge-to-total light ratio;
Column (6): BH mass derived using the $M_{\mathrm{BH}}$--$\sigma_\ast$ relation from equation (1), or if noted, direct BH mass measurement;
Column (7): Eddington ratio.\\
$^{a}$ Direct $M_{\mathrm{BH}}$ measurement from gas kinematics \citep{atkinson05}.\\
$^{b}$ Velocity dispersion of the nuclear star cluster.\\
$^{c}$ Direct $M_{\mathrm{BH}}$ measurement from gas kinematics \citep{sarzi01}.\\
$^{d}$ Direct $M_{\mathrm{BH}}$ measurement from stellar kinematics \citep{nowak10}.
}
\end{deluxetable*}

%%%%%%%%%%%%%%%%%%%%%%%%
\begin{figure}[t]
\centering
\resizebox{0.85\hsize}{!}{\includegraphics{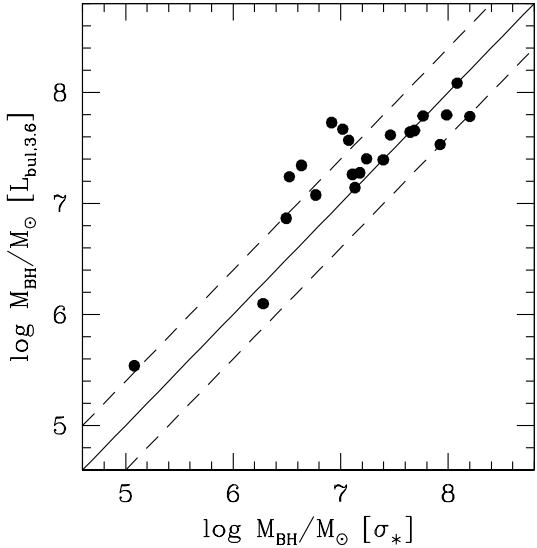}}
\caption{Comparison between black hole masses obtained from bulge luminosities against those obtained from central stellar velocity dispersions. The solid line shows the exact correspondence between both methods, and the dashed lines are offset $\pm$0.4 dex, indicating the typical intrinsic scatter in these scaling relations.\label{figmbh}}
\end{figure}
%%%%%%%%%%%%%%%%%%%%%%%%

\subsection{Eddington Ratios}

While the X-ray luminosity attributed to the BH feeding process is a good proxy of the degree of nuclear activity, a clearer picture of the actual level of accretion will come from the Eddington ratio ($L_{\mathrm{bol}}/L_{\mathrm{Edd}}$), where $L_{\mathrm{bol}}$ and $L_{\mathrm{Edd}}$ correspond to the bolometric and Eddington luminosities respectively.
The X-ray bolometric correction, $L_{\mathrm{bol}}/L_{\mathrm{X}}$, has been found to depend strongly on Eddington ratio \citep{vasudevan&fabian07}, with LLAGNs and their modest accretion rates requiring lower correction values when compared to higher-luminosity AGNs and quasars.
While the Eddington ratios of LLAGNs span over six orders of magnitude ($10^{-8}$--$10^{-2}$) and adopting a single bolometric correction might seem too simplistic, the strongest dependence between these two quantities starts above $L_{\mathrm{bol}}/L_{\mathrm{Edd}}\sim 10^{-1}$, below which the behavior of the bolometric correction is rather flat.
Therefore, we use $L_{\mathrm{bol}}/L_{\mathrm{X}}$=15.8 from \citet{ho09b}, derived from a sample of nearby LLAGNs with robust spectral energy distributions.

The Eddington luminosity will normalize the X-ray luminosities by BH mass and it is defined as
$L_{\mathrm{Edd}} = 1.26 \times 10^{38}\, (M_{\mathrm{BH}}/M_{\odot})$ erg s$^{-1}$.
Only four galaxies from our sample (NGC 1300, NGC 2787, NGC 3368, and NGC 4596) have direct $M_{\mathrm{BH}}$ measurements derived either from stellar or gas kinematics.
For the remaining galaxies of the sample, it is possible to predict their BH masses through the empirical scaling relations between $M_{\mathrm{BH}}$ and galaxy properties such as bulge luminosity \citep{mbh_l1, mbh_l2}, central stellar velocity dispersion \citep{mbh_sigma1, mbh_sigma2, mbh_sigma3}, and bulge stellar mass \citep{mbh_m1, mbh_m2}.
Among these, the correlation between BH mass and central stellar velocity dispersion ($\sigma_\ast$) has been found to be the most significant one \citep{gebhardt03}.
Twenty-six of the remaining galaxies have available $\sigma_\ast$ measurements from the literature, allowing us to apply the $M_{\mathrm{BH}}$--$\sigma_\ast$ relation under the assumption that it holds true for the galaxies probed in the present study\footnote{We caution that it is still under debate whether the $M_{\mathrm{BH}}$--$\sigma_\ast$ holds universally other than for classical bulges and elliptical galaxies \citep[e.g.,][]{kormendy&ho13}.}.
Since the establishment of this scaling relation, the number of  $M_{\mathrm{BH}}$ measurements has been substantially expanded, and hence we opt to use the updated relation from \citet{gultekin09}, given by:

\begin{equation}
\mathrm{log}\,(M_{\mathrm{BH}}/M_{\odot}) = 8.12 + 4.24\, \mathrm{log} \left( \frac{\sigma_{\ast}}{ 200\, \mathrm{km}\, \mathrm{s}^{-1} } \right).
\end{equation}

For the rest of our sample lacking $\sigma_\ast$ measurements, BH masses can be predicted by employing the correlation with bulge luminosity at 3.6 $\mu$m ($L_{bul,3.6}$) obtained by \citet{sani11}:

\begin{equation}
\mathrm{log}\,(M_{\mathrm{BH}}/M_{\odot}) = 8.19 + 0.93\, \mathrm{log} \left( \frac{ L_{bul,3.6}}{ 10^{11} L_{\odot ,3.6} } \right).
\end{equation}

Bulge luminosities in solar units at 3.6 $\mu$m are derived using the bulge-to-total light ratios ($B/T$) from the two-dimensional image decomposition (see section 4.1 for details) together with the 3.6 $\mu$m absolute magnitude of the galaxies \citep{munoz-mateos13}.
To estimate luminosities in solar units, we use the 3.6 $\mu$m solar absolute magnitude value of $M_{\odot}^{3.6}=3.24$ derived by \citet{oh08}.
Within these galaxies, six were best-modeled without a bulge component ($B/T=0$), making it not possible to estimate their $M_{\mathrm{BH}}$ through this method, yet all of them had class IV nuclei, i.e., no X-ray nuclear source.

Both $M_{\mathrm{BH}}$--$\sigma_\ast$ and  $M_{\mathrm{BH}}$--$L_{bul,3.6}$ relations used here have an intrinsic scatter of $\sim$0.4 dex, being the dominant source of uncertainty in our $M_{\mathrm{BH}}$ estimates.
We assess the consistency of both methods at predicting BH masses through a direct comparison:
in Figure \ref{figmbh}, we plot $M_{\mathrm{BH}}$ estimates using both methods on 22 galaxies with $\sigma_\ast$ measurements as well as $B/T>0$.
In the figure, the solid line indicates the exact correspondence between both methods, and the dashed lines mark the 0.4 dex intrinsic scatter from the scaling relations.
The bulk of the galaxies obey the relation within the uncertainties, and $M_{\mathrm{BH}}$ derived from bulge luminosities are, on average, $\sim$0.1 dex higher than those from $\sigma_\ast$. 
This is particularly interesting, as it has been argued that barred galaxies appear systematically offset $\sim$-0.5 dex from the $M_{\mathrm{BH}}$--$\sigma_\ast$ relation \citep{graham08,gadotti09} and hence these BH masses could be overestimated.
At least for the galaxies probed here, this effect is not observed, and both methods can be considered consistent within the scatter.
The resulting BH masses and corresponding Eddington ratios, along with the relevant parameters used for their calculation, are presented Table 2.

\section{Quantifying bar strength}

Stellar bars come in very different shapes and sizes, and therefore it would be unfair to simply categorize them as a single group.
Bars have different strengths which will determine how efficiently they can drive the interstellar medium to central regions of the galaxy.
As bars evolve with time, their pattern speeds slow down, allowing them to become more elongated and eccentric, and therefore stronger \citep[for a review, see][]{athanassoula12}.
While the pattern speed of bars is difficult to measure, their structure can be quantified from two-dimensional image modeling of the galaxy components.
A different approach comes from quantifying the gravitational torques due to non-axisymmetric structures \citep[e.g.,][]{stark77, combes&sanders81, zaritsky86}.
Given that the NIR imaging mostly probes old stars (and hence stellar mass), one can infer the gravitational potential from the bar without the need of defining its structure.
Below we describe and apply these two independent approaches at quantifying bar strength: one based on its structure and the other on its gravitational potential.

%%%%%%%%%%%%%%%%%%%%%%%
\begin{deluxetable}{lr@{ $\pm$ }lr@{ $\pm$ }lc}
\tabletypesize{\scriptsize}
\tablecaption{Bar structural properties and maximum relative torque \label{tab3}}
\tablehead{Galaxy & \multicolumn{2}{c}{$\epsilon$} & \multicolumn{2}{c}{$c$} & $Q_b$ \\
(1) &  \multicolumn{2}{c}{(2)} &  \multicolumn{2}{c}{(3)} &  (4)
}
\startdata
NGC 255& 0.60 & 0.02 &  2.64 & 0.18 &  $0.51^{+0.05}_{-0.04}$ \\
NGC 685& 0.63 & 0.02 &  2.75 & 0.61 &  $0.39^{+0.04}_{-0.03}$ \\
NGC 1036& 0.37 & 0.01 &  2.66 & 0.20 &  $0.34^{+0.03}_{-0.02}$ \\
NGC 1073& 0.72 & 0.01 &  2.78 & 0.55 &  $0.63^{+0.07}_{-0.08}$ \\
NGC 1097& 0.45 & 0.09 &  2.77 & 1.19 &  $0.26^{+0.04}_{-0.04}$ \\
NGC 1232& 0.35 & 0.01 &  2.76 & 1.07 &  $0.13^{+0.01}_{-0.01}$ \\
NGC 1291& 0.64 & 0.01 &  2.78 & 0.01 &  $0.14^{+0.02}_{-0.02}$ \\
NGC 1300& 0.75 & 0.01 &  3.16 & 0.20 &  $0.57^{+0.12}_{-0.10}$ \\
NGC 1302& 0.48 & 0.01 &  2.81 & 0.01 &  $0.10^{+0.01}_{-0.01}$ \\
NGC 1341& 0.61 & 0.01 &  3.00 & 0.51 &  $0.51^{+0.04}_{-0.05}$ \\
NGC 1367& 0.54 & 0.01 &  2.72 & 0.28 &  $0.13^{+0.02}_{-0.02}$ \\
NGC 1493& 0.63 & 0.03 &  2.64 & 0.75 &  $0.41^{+0.06}_{-0.04}$ \\
NGC 1637& 0.65 & 0.01 &  2.73 & 0.12 &  $0.23^{+0.04}_{-0.03}$ \\
NGC 1640& 0.65 & 0.01 &  2.92 & 0.28 &  $0.28^{+0.05}_{-0.04}$ \\
NGC 1672& 0.63 & 0.09 &  3.90 & 2.17 &  $0.37^{+0.06}_{-0.06}$ \\
NGC 2787& 0.69 & 0.01 &  2.77 & 0.32 &  $0.15^{+0.02}_{-0.02}$ \\
NGC 3344& 0.46 & 0.03 &  2.63 & 0.62 &  $0.06^{+0.01}_{-0.01}$ \\
NGC 3351& 0.70 & 0.02 &  2.95 & 0.33 &  $0.24^{+0.04}_{-0.04}$ \\
NGC 3368& 0.51 & 0.01 &  2.51 & 0.29 &  $0.24^{+0.03}_{-0.03}$ \\
NGC 3627& 0.67 & 0.05 &  2.99 & 1.14 &  $0.31^{+0.09}_{-0.06}$ \\
NGC 4136& 0.68 & 0.05 &  2.95 & 0.77 &  $0.11^{+0.03}_{-0.02}$ \\
NGC 4245& 0.62 & 0.02 &  2.83 & 0.20 &  $0.19^{+0.03}_{-0.02}$ \\
NGC 4303& 0.57 & 0.02 &  3.12 & 0.17 &  $0.42^{+0.08}_{-0.08}$ \\
NGC 4314& 0.75 & 0.01 &  2.76 & 0.13 &  $0.45^{+0.08}_{-0.08}$ \\
NGC 4394& 0.62 & 0.02 &  2.85 & 0.45 &  $0.23^{+0.04}_{-0.03}$ \\
NGC 4450& 0.34 & 0.01 &  3.10 & 0.15 &  $0.14^{+0.02}_{-0.02}$ \\
NGC 4548& 0.68 & 0.06 &  2.87 & 0.97 &  $0.28^{+0.04}_{-0.04}$ \\
NGC 4579& 0.49 & 0.10 &  2.68 & 1.26 &  $0.18^{+0.03}_{-0.03}$ \\
NGC 4596& 0.68 & 0.01 &  2.78 & 0.41 &  $0.25^{+0.05}_{-0.04}$ \\
NGC 4639& 0.60 & 0.01 &  2.94 & 0.11 &  $0.27^{+0.03}_{-0.04}$ \\
NGC 4713& 0.15 & 0.04 &  2.78 & 0.48 &  $0.25^{+0.04}_{-0.04}$ \\
NGC 4725& 0.54 & 0.09 &  3.00 & 1.74 &  $0.24^{+0.03}_{-0.03}$ \\
NGC 5350& 0.70 & 0.01 &  2.71 & 0.33 &  $0.44^{+0.08}_{-0.07}$ \\
NGC 5371& 0.58 & 0.05 &  2.97 & 0.24 &  $0.13^{+0.03}_{-0.02}$ \\
NGC 5584& 0.61 & 0.01 &  2.71 & 0.25 &  $0.40^{+0.03}_{-0.03}$ \\
NGC 5728& 0.51 & 0.02 &  2.93 & 0.52 &  $0.41^{+0.05}_{-0.06}$ \\
NGC 5964& 0.55 & 0.01 &  3.04 & 0.38 &  $0.94^{+0.16}_{-0.17}$ \\
NGC 7479& 0.68 & 0.01 &  2.80 & 0.43 &  $0.54^{+0.11}_{-0.10}$ \\
NGC 7552& 0.64 & 0.01 &  2.60 & 0.07 &  $0.36^{+0.08}_{-0.07}$ \\
NGC 7743& 0.54 & 0.04 &  2.68 & 0.81 &  $0.19^{+0.02}_{-0.02}$ \\
PGC 3853& 0.63 & 0.04 &  2.78 & 0.68 &  $0.45^{+0.05}_{-0.05}$
\enddata
\tablecomments{Column (1): galaxy name;
Column (2): deprojected bar ellipticity;
Column (3): bar boxiness; 
Column (4): maximum relative torque.
}
\end{deluxetable}
%%%%%%%%%%%%%%%%%%%%%%%%

\subsection{Bar Structure: Ellipticity and Boxiness}

Suggested early on by analytical models as a fundamental parameter describing a barred galaxy and its dynamical evolution \citep{athanassoula92a}, the deprojected bar ellipticity was proposed by \citet{martin95} as quantifiable measure of bar strength, in the sense that the smaller the axial ratio, the stronger the non-axisymmetric force the bar will be able to exert.
Interestingly, Martin found that for a small sample of galaxies with nuclear starbursts, the majority were hosted by galaxies with highly eccentric bars.
Bar ellipticity has been widely used in the literature \citep[e.g.,][]{rozas98, abraham99, aguerri99, knapen00, shlosman00, laine02, gadotti11,wang12} and has the advantage of being readily available from photometric images and is independent on assumptions of the galaxy's physical properties.
Additionally, N-body simulations have shown that as a bar grows stronger, it does not only get more eccentric but also more boxy in shape \citep{athanassoula02a}.
This was observed by \citet{gadotti11} who found that bar ellipticity and boxiness were correlated, and defined their product as a proxy for bar strength.

In this paper, we use the structural parameters derived for the parent sample of S$^{4}$G barred galaxies to be presented in detail in T.~Kim et al. (in preparation).
The two-dimensional image decomposition code {\ttfamily BUDDA} \citep{budda, gadotti08} was used to model the galaxies:
three components, described by concentric ellipses, were used to represent the bulge, disk, and bar.
When necessary, a central point source component was included to account for a bright AGN or nuclear star cluster.
In some cases, nuclear rings were masked and disk breaks, i.e., disk light profiles with two slopes, were accounted for.
The careful procedures adopted assure that the structural properties of the bar are always accurately measured.
The structural parameters of interest are derived from the equation of a generalized ellipse \citep{athanassoula90}:

\begin{equation}
\left(\frac{|x|}{a} \right)^c + \left(\frac{|y|}{b} \right)^c = 1,
\end{equation}
where $x$ and $y$ are the pixel coordinates, $a$ and $b$ are the semimajor- and semiminor-axes respectively,
and the exponent $c$ describes the bar boxiness: if $c>2$, the bar is boxy, $c<2$ the bar is disky, and if $c=2$ the shape of bar is a perfect ellipse.
The observed ellipticity of the bar, defined as $\epsilon_o=1-b/a$, is deprojected following the analytical expressions from \citet{gadotti07}.
The boxiness parameter is kept fixed at $c=2$ for the bulge  and disk components, which are thus always described using perfect ellipses.
In terms of surface brightness, we model the disk with an exponential profile \citep{freeman70} allowing the disk to have a break.
Both bulge and bar were modeled using a \citet{sersic} profile.

The resulting deprojected ellipticity and boxiness measurements are presented in Table \ref{tab3}.

\subsection{Gravitational Torques}

Proposed by \citet{combes&sanders81}, a more sophisticated approach at measuring bar strengths comes from directly estimating tangential forces in 
the bar region and comparing them to the axisymmetric potential of the disk.
This force ratio represents a measure of the bar-induced gravitational torque, and it is defined as:

\begin{equation}
Q_{T} (r) = \frac{F_{T}^{\mathrm{max}} (r)}{\langle F_{R} (r)\rangle},
\end{equation}
where, at a given radius $r$, $F_{T}^{\mathrm{max}} (r)$ corresponds to the maximum amplitude of the tangential force
and $\langle F_{R} (r)\rangle$ is the mean axisymmetric radial force at that radial distance.
The force ratio parameter $Q_{T}$ varies with radius, and in order to implement a single measure of bar strength for the whole galaxy, $Q_{b}$ is adopted as the maximum value of $Q_{T}$ at the bar region.
Based on the practical implementation of the gravitational torque method by \citet{quillen94}, \citet{buta&block01} measured the force ratio parameter $Q_{b}$ for 36 nearby spiral galaxies from NIR images.
They found that galaxies categorized from their apparent bar strength through the \citet{devaucouleurs59a} classification scheme could have a wide range of {\em true} bar strengths.
From hydrodynamic simulations, it has been shown that $Q_{b}$ is directly related to the bar-driven mass inflow rate \citep{kim12b}, making it a highly relevant parameter when studying the impact of bar strength on nuclear activity.

For this study, we use the $Q_{b}$ bar strength measurements to be presented in detail by Sim\'on D\'iaz-Garc\'ia et al. (in preparation), who compute non-axisymmetric forces on an extended sample of S$^4$G spiral galaxies.
The calculations are performed with the polar grid method, also accounting for artificial bulge stretch due to deprojection \citep[see][]{salo10}.
Gravitational potentials were inferred under two main assumptions:
(1) 3.6 $\mu$m light traces stellar mass with a constant mass-to-light ratio\footnote{This assumption has been shown to be fairly reasonable \citep{eskew12}, although see \citet{meidt12} for a careful treatment of the separation of old stellar light in 3.6 $\mu$m images from the emission of polycyclic aromatic hydrocarbons, hot dust, and young stars.},
and (2) the vertical scale height of the disk, $h_{z}$, scales with the disk size as $h_{z}=0.1\,r_{\mathrm{K20}}$ \citep{speltincx08}, where $r_{\mathrm{K20}}$ is the K-band surface brightness isophote of 20 mag arcsec$^{-2}$ from 2MASS.
For further technical details on the method, see, e.g., \citet{buta04,laurikainen04a, laurikainen04b}.
The resulting $Q_b$ measurements for our sample are presented in Table \ref{tab3}.

\section{Results}

\subsection{Nuclear X-ray Sources}

Out of the 41 galaxies analyzed, we detected X-ray nuclear point-like sources (classes I and II as described in Section 3) in 31 of them.
Within these, nine have been previously classified as Seyferts, six as low-ionization nuclear emission line regions (LINERs), and one as an H~\textsc{ii} nucleus, as indicated in Table \ref{tab1}.
None of the galaxies without a nuclear detection (classes III and IV) has been previously classified as active based on optical diagnostics.

The distributions of X-ray luminosities and Eddington ratios are shown in the top panels of Figure \ref{figstr}.
For both quantities, our sample as a whole spans around six orders of magnitude in agreement with previous studies of X-ray nuclear activity in nearby galaxies \citep[e.g.,][]{zhang09,ho09b}, with a median $L_{\mathrm{X}}$ = $2.6 \times 10^{38}$ erg s$^{-1}$ and a median $L_{\mathrm{bol}}$/$L_{\mathrm{Edd}}$ = $5.4 \times 10^{-6}$.
If we consider only those galaxies with nuclear detections (classes I and II), the median values are $L_{\mathrm{X}}$ = $4.3 \times 10^{38}$ erg s$^{-1}$ and $L_{\mathrm{bol}}$/$L_{\mathrm{Edd}}$ = $6.9 \times 10^{-6}$, consistent with the median values of the AGN sample from the Palomar Survey reported by \citet{ho09b}.

A caveat concerning X-ray studies of low-luminosity AGNs lies in the possibility that these nuclear X-ray point-sources could not necessarily be accreting BHs.
Possible confusion with other sources such as low-mass X-ray binaries (LMXBs) has been discussed extensively in the literature, and different arguments have been invoked in favor of the AGN nature of nuclear point-like sources coincident with the independently determined center \citep[e.g.,][]{gallo08, desroches&ho09, zhang09, grier11, jenkins11}.
Perhaps one of the most compelling arguments comes from the probability to get an X-ray binary precisely at the nuclear position:
based on the LMXB population study by \citet{gilfanov04}, the analyses by \citet{gallo08} and \citet{zhang09} estimate of the order of 10$^{-2}$ LMXBs brighter than $\sim$10$^{38}$ erg s$^{-1}$ within an aperture of the size of the {\em Chandra} PSF.
Together with the excellent agreement between the nuclear and NIR position, comparable to their astrometric uncertainties of $\sim$1\arcsec, as well as the lack of other point-like X-ray sources in the immediate vicinity ($\sim$5\arcsec) for the vast majority of our sample, the existence of any significant contamination from LMXBs can be likely ruled out.

\subsection{Bar Strength Versus Nuclear Activity}

\begin{figure}[t]
\centering
\resizebox{1\hsize}{!}{\includegraphics{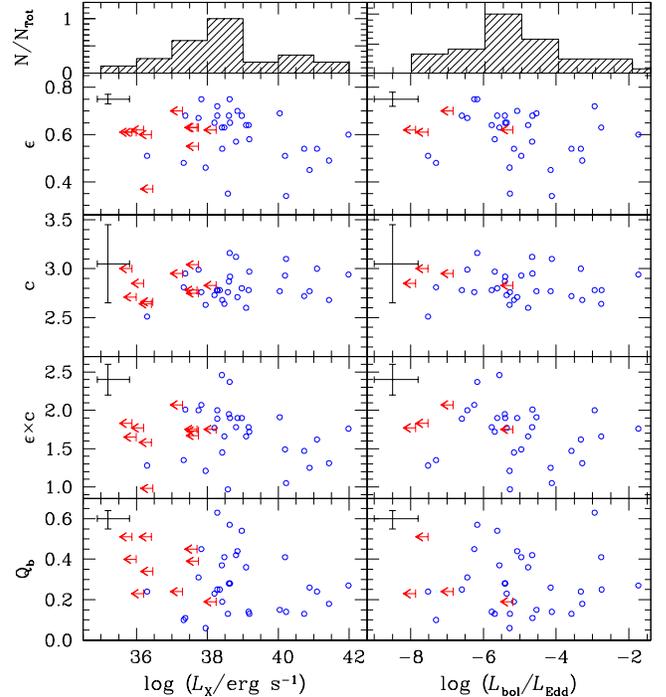}}
\caption{The top panels show the distributions of 2--10 keV X-ray luminosities (left) and Eddington ratios (right).
Below, from top to bottom, deprojected ellipticity $\epsilon$, boxiness $c$, their product $\epsilon \times c$, and gravitational torque parameter $Q_b$, are plotted against $L_{\mathrm{X}}$ and $L_{\mathrm{bol}}$/$L_{\mathrm{Edd}}$.
The arrows represent upper limits, i.e., nuclear classes III and IV.
On the top left corner of each panel, we show the mean measurement uncertainties.
\label{figstr}}
\end{figure}

\begin{figure}[t]
\centering
\resizebox{1\hsize}{!}{\includegraphics{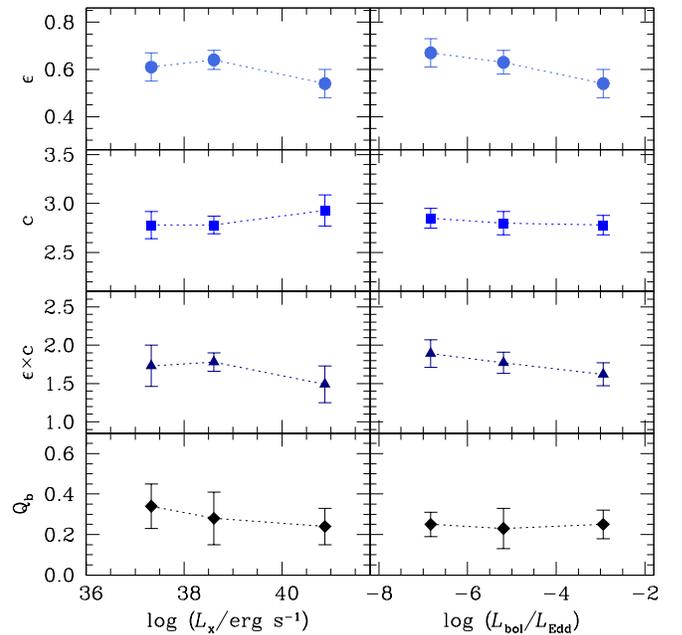}}
\caption{Median bar strength of our sample as a function of AGN activity, binned in X-ray luminosity (left) and Eddington ratio (right), with boundaries set at $L_{\mathrm{X}}$ = 10$^{38}$, 10$^{40}$ erg~s$^{-1}$ and $L_{\mathrm{bol}}$/$L_{\mathrm{Edd}}$ = 10$^{-6}$, 10$^{-4}$ respectively, and centered at the median values of their respective bin.
From top to bottom, the median of deprojected ellipticity, boxiness, their product $\epsilon \times c$, and gravitational torque, are plotted following the Y-axis ranges as in Figure \ref{figstr}.
Vertical error bars correspond to the median absolute deviations.
\label{figstr2}}
\end{figure}

In Figure \ref{figstr}, we plot the different measurements of bar strength against nuclear activity as described in the previous sections.
Deprojected ellipticity $\epsilon$, boxiness $c$, their product $\epsilon \times c$, and gravitational torque parameter $Q_b$ are shown from top to bottom;
on the left-side panels against 2--10 keV X-ray luminosity $L_{\mathrm{X}}$, and on the right-side panels against Eddington ratio $L_{\mathrm{bol}}$/$L_{\mathrm{Edd}}$.
One can immediately observe that no clear correlations are present:
for any given luminosity or Eddington ratio there is a wide range of possible bar strengths, as quantified through the four methods presented before.
To better explore possible trends in our sample, a more practical representation of our results is shown in Figure \ref{figstr2}.
Here, we plot the median of each bar strength indicator versus AGN activity, binning our sample in both X-ray luminosity and Eddington ratio with boundaries at $L_{\mathrm{X}}$ = 10$^{38}$, 10$^{40}$ erg~s$^{-1}$ and $L_{\mathrm{bol}}$/$L_{\mathrm{Edd}}$ = 10$^{-6}$, 10$^{-4}$ respectively.
In this plot, the lack of any relevant positive trend between bar strength and AGN activity is even more evident.
On the contrary, in a few cases there seems to be a negative trend of bar strength with increasing AGN luminosity or accretion rate, yet the large dispersions cast doubt on their significance.

%%%%%%%%%%%
%%%%%%%%%%%

%%%%%%%%%%%
%%%%%%%%%%%

%%%%%%%%%%%%%%%%%%%%%%%
\begin{deluxetable*}{lcccccccccccc}
\tabletypesize{\scriptsize}
\tablecaption{Correlation tests\label{tab4}}
\tablehead{ & & \multicolumn{2}{c}{All galaxies} & & \multicolumn{2}{c}{Classes I+II} & & \multicolumn{2}{c}{``Faint'' ($M^{3.6}$ $>$ -20.82)} & & \multicolumn{2}{c}{``Bright'' ($M^{3.6}$ $<$ -20.82)} \\ \cline{3-4}\cline{6-7}\cline{9-10}\cline{12-13}
 & & $\rho$ & Significance & & $\rho$ & Significance & & $\rho$ & Significance & & $\rho$ & Significance
}
\startdata
$L_{\mathrm{X}}$ vs \\
& $\epsilon$ &    -0.11 &     0.51 & &    -0.20 &     0.28 & &     0.16 &     0.50 & &    -0.36 &     0.11 \\ 
& $c$ &     0.12 &     0.44 & &     0.14 &     0.45 & &    -0.02 &     0.93 & &    -0.08 &     0.72 \\ 
& $\epsilon \times c$ &    -0.10 &     0.54 & &    -0.21 &     0.25 & &     0.09 &     0.72 & &    -0.40 &     0.07 \\ 
& $Q_b$ &    -0.19 &     0.24 & &     0.01 &     0.97 & &    -0.31 &     0.18 & &    -0.09 &     0.70 \\ \hline
$L_{\mathrm{bol}}/L_{\mathrm{Edd}}$ vs \\
& $\epsilon$ &    -0.28 &     0.11 & &    -0.28 &     0.13 & &    -0.15 &     0.60 & &    -0.42 &     0.06 \\ 
& $c$ &    -0.17 &     0.33 & &    -0.08 &     0.65 & &    -0.13 &     0.67 & &    -0.06 &     0.78 \\ 
& $\epsilon \times c$ &    -0.34 &     0.04 & &    -0.31 &     0.09 & &    -0.30 &     0.30 & &    -0.40 &     0.07 \\ \smallskip
& $Q_b$ &    -0.03 &     0.87 & &     0.00 &     1.00 & &    -0.04 &     0.89 & &     0.00 &     0.98\\  \hline \\
%%%%%%%%%%%%%%%%%%%%%%%
 & & \multicolumn{2}{c}{``Early'' ($T$-type $<$ 3.1)} & & \multicolumn{2}{c}{``Late'' ($T$-type $>$ 3.1)} & & \multicolumn{2}{c}{``Bulgy'' ($B/T$ $>$ 0.13)} & & \multicolumn{2}{c}{``Disky'' ($B/T$ $<$ 0.13)} \\ \cline{3-4}\cline{6-7}\cline{9-10}\cline{12-13}
  & & $\rho$ & Significance & & $\rho$ & Significance & & $\rho$ & Significance & & $\rho$ & Significance\\ \hline
$L_{\mathrm{X}}$ vs \\
& $\epsilon$ &    -0.37 &     0.10 & &     0.15 &     0.51 & &    -0.35 &     0.14 & &     0.05 &     0.82 \\ 
& $c$ &     0.09 &     0.71 & &     0.21 &     0.37 & &     0.15 &     0.52 & &     0.14 &     0.54 \\ 
& $\epsilon \times c$ &    -0.34 &     0.13 & &     0.17 &     0.48 & &    -0.32 &     0.16 & &     0.06 &     0.79 \\ 
& $Q_b$ &    -0.06 &     0.79 & &    -0.02 &     0.94 & &     0.00 &     0.99 & &    -0.11 &     0.62 \\ \hline
$L_{\mathrm{bol}}/L_{\mathrm{Edd}}$ vs \\
& $\epsilon$ &    -0.36 &     0.11 & &    -0.17 &     0.55 & &    -0.35 &     0.13 & &    -0.23 &     0.40 \\ 
& $c$ &     0.04 &     0.85 & &    -0.48 &     0.08 & &     0.12 &     0.63 & &    -0.50 &     0.06 \\ 
& $\epsilon \times c$ &    -0.36 &     0.11 & &    -0.31 &     0.27 & &    -0.35 &     0.13 & &    -0.36 &   0.18 \\ 
& $Q_b$ &    -0.08 &     0.74 & &    -0.12 &     0.69 & &     0.00 &     0.99 & &    -0.14 &     0.62
\enddata
\tablecomments{Spearman's rank correlation coefficient $\rho$ and its significance are measured for the whole sample as well as for subsamples excluding upper limits, and divided according to the median 3.6 $\mu$m absolute magnitude, morphological $T$-type, and bulge-to-total light ratio of the sample: $M^{3.6}$$=$-20.82 (or  $M_{\ast}$ $\sim$ $2.8\times10^{10}$$M_{\odot}$), $T$-type = 3.1, and $B/T$ = 0.13 respectively.
When a perfect correlation (or anticorrelation) occurs, $\rho$ adopts 1 (or -1), whereas $\rho=0$ if no correlation is present.
The significance of the correlation is a value within [0,1], and should be consistent with zero in case of a significant correlation.
}
\end{deluxetable*}
%%%%%%%%%%%%%%%%%%%%%%%

To quantitatively investigate whether any of these trends are significant and could reveal a link between bar strength and nuclear activity, we test the dependence between X-ray luminosity and Eddington ratio against the different bar strength proxies using Spearman's rank correlation.
Besides performing this statistical test for our entire sample, we also test for correlations excluding galaxies with nuclear classes III or IV:
even though we are operating under the assumption that all of these galaxies have a BH at their centers, we account for the possibility that the lack of detection in those cases for which we present upper limits may be due to the lack of a BH.
In such a case, these galaxies should not be part of the correlation and might be affecting the overall result.
Therefore, we test this possibility by performing the correlation test only on those galaxies with nuclear detections (classes I and II).
Additionally, we test whether any correlation shows up when probing different subsamples based on luminosity, morphological $T$-type, and bulge-to-total light ratio cuts at their respective median values.
The effect of bar strength on nuclear activity in different galaxy luminosity --and hence stellar mass-- regimes is tested by defining ``Faint'' and ``Bright'' subsamples based on a luminosity cut at $M^{3.6}$=-20.82, which corresponds to $M_{\ast}$ $\sim$ $2.8\times10^{10}$$M_{\odot}$ following the conversion presented in the Appendix of \citet{munoz-mateos13};
``Early'' and ``Late'' morphological subsamples are defined dividing at $T$-type = 3.1;
and ``Bulgy'' and ``Disky''  subsamples based on the bulge-to-total light ratio from the image decomposition are defined by a cut at $B/T$ = 0.13.

In Table \ref{tab4}, we show the Spearman's rank correlation coefficient $\rho$ and its significance for every combination of bar strength and AGN activity measurement for the eight samples analyzed, i.e., whole sample, galaxies with nuclear X-ray point source, and subsamples divided by 3.6 $\mu$m absolute magnitude, morphological $T$-type, and bulge-to-total light ratio.
No significant correlation is obtained in any of the subsamples.
The Spearman's coefficient and significance values for the whole sample reflect the trends shown in Figure \ref{figstr2}, and no major changes happen by excluding upper limits from the analysis.
Dividing the sample of galaxies by their 3.6 $\mu$m absolute magnitude does not particularly change the trends, yet it strengthens the, albeit still not significant, anticorrelation between $\epsilon \times c$ and AGN activity as traced by both X-ray luminosity and Eddington ratio for the more luminous, massive end of our sample.
As one could expect from the correspondence between Hubble type and bulge extent relative to the galaxy, both pairs of subsamples, ``Early''/``Bulgy'' and ``Late''/``Disky'' show a very good agreement in terms of their correlation scores.
However, no significant correlation shows up in neither of them, suggesting that the presence or absence of a significant bulge component does not affect the influence of the stellar bar on the nuclear fueling.
This is particularly interesting for the case of $Q_b$, which is directly affected by the bulge:
the relative torque parameter is diluted in the presence of a stronger axisymmetric component, i.e., a more massive bulge.
Therefore, if there was a direct connection between the non-axisymmetric gravitational potential from the stellar bar and the level of AGN activity, one would have expected to see it at least in the subsample of galaxies with less massive bulges.

In summary, no significant correlation is found for any of the subsamples probed, and therefore our analysis indicates an independence between bar strength and degree of nuclear activity irrespective of galaxy luminosity, stellar mass, morphology, or bulge relative size.

\section{Discussion}

From the point of view of both, simulations and observations, stellar bars have been shown to be able to drive material toward the central regions of a galaxy.
The notion that bars are also able to feed a BH however, has not been supported by empirical results.
Most studies investigating whether bars had any impact on AGN activity did so by measuring bar fractions among samples of active and inactive galaxies, or alternatively, by measuring the AGN fraction between barred and unbarred galaxies.
Since bars have a wide range of strengths, and AGN activity has a continuous distribution in luminosity and mass accretion rate spanning a few orders of magnitude, perhaps most previous attempts at connecting bars and AGN oversimplified on their approach by discretizing these quantities.
Among the few studies taking this into account, \citet{ho97b} investigated AGN luminosity distributions, as traced by the nuclear H$\alpha$ emission, on barred and unbarred galaxies finding that the presence of a bar had no influence on the observed nuclear luminosity.
On the other hand, \citet{laurikainen02a, laurikainen04a} quantified bar strengths using the gravitational torque parameter $Q_b$ for samples of barred active and non-active galaxies and found no evidence that would suggest that stronger bars, as traced by $Q_b$, tend to favor AGN host galaxies.
In fact, they found weaker $Q_b$ values among active galaxies against their inactive counterparts, yet they highlight that this is a side-effect of $Q_b$ being tied to Hubble type,
in the sense that a more massive bulge relative to the disk will induce a stronger axisymmetric potential, washing out the bar-induced torque.
Therefore, early-type spirals, where the optically classified AGNs analyzed in these studies were preferentially found, will have intrinsically weak bars according to $Q_b$.
The inverse effect was observed by \citet{laurikainen04a} when comparing $m=2$ Fourier amplitude of density of the bar, in the sense that early-type spirals have larger values when compared to later-types, in which inactive galaxies were mostly found.
Both effects, however, go away if Hubble-type is kept fixed, with active and inactive galaxies showing comparable values of these bar strength indices.

In the context of BH accretion rates, \citet{crenshaw03} compared the fraction of bars between two subclasses of active galaxies:
narrow-line and broad-line Seyfert 1s (NLS1s and BLS1s respectively).
At a fixed luminosity, the former have lower-mass BHs compared to the latter, and given their near-Eddington accretion rates, NLS1s are thought to be AGNs in an early stage of their activity \citep{mathur00}.
Additionally, NLS1s tend to host pseudo-bulges \citep{orbandexivry11, mathur12}, making them ideal test-cases for the study of secular processes driving the evolution of galaxy and BH.
Crenshaw et al. found that bars are indeed more frequent in NLS1s, suggesting a scenario in which their higher accretion rates are related to the bar-induced fueling.
Other studies have tackled the impact of bars on the Eddington ratio using large samples of galaxies from the Sloan Digital Sky Survey \citep{sdss0} with mixed results:
while \citet{alonso13} argue that barred active galaxies show higher mean accretion rates against their unbarred counterparts, \citet{lee12} found that both barred and unbarred active galaxies have consistent Eddington ratio distributions. 

A connection between AGN activity and host galaxy on kiloparsec scales has been pursued not only from the point of view of stellar bars, but also from the perspective of the kinematics of the galaxy.
On a comparison between the stellar and gaseous kinematics within the central kiloparsec of Seyfert and inactive galaxies, \citet{dumas07} found no remarkable differences on large scales, with both stars and gas showing regular rotation patterns and a general alignment with each other.
On smaller scales however, within the inner few hundred parsecs, the ionized gaseous component of active galaxies is more disturbed compared to their inactive counterparts, leading to the reasonable conclusion that signatures of the ongoing BH feeding can only be found in the innermost regions of the galaxy.

In this respect, {\it HST} programs have targeted the nuclear regions of active galaxies to study their nuclear dust structure \citep[e.g.,][]{regan99a, martini99}.
The morphology of the circumnuclear dust can reveal whether the influence of the bar extends to the unresolved nucleus in the shape of straight dust lanes.
Surprisingly, these studies found these signatures only in a minority of active galaxies, and found that another observed mechanism, nuclear dust spirals, might be responsible of driving the gas further down to parsec scales.
Nevertheless, comparisons between the circumnuclear dust structure of active and inactive galaxies have shown that nuclear dust spirals are equally common on both samples, without a preference for active nuclei \citep{martini03b}, hinting at the possibility that the lifetime of AGN activity has to be less than the inflow time from these structures.
Furthermore, there is no correlation between the structure of the circumnuclear dust and the strength of the stellar bar: 
\citet{peeples06} found that strongly barred galaxies can have a wealth of nuclear dust morphologies, ranging from a clearly defined nuclear dust spiral to a chaotic structure unlikely to be able to drive material to the very central regions, suggesting that a strong bar does not necessarily imply an efficient nuclear fueling.

\subsection{On the Stability of Bars}

Based on the results presented here, nuclear luminosity and BH accretion rate are not influenced by the strength of the large-scale bar.
Do our findings imply that bars play no role in driving the gas that would eventually fuel an AGN?
The only safe conclusion one can draw from our results is that the {\em current} strength of the stellar bar has no impact on the level of co-occurrent AGN activity, and hence, if bars were to weaken over time while driving gas down to the galactic centers, we could be missing its true influence on nuclear activity.

Early simulations of the dynamical evolution of bars in galaxies suggested that bar-induced gas inflows initiate the growth of a central ($r \lesssim$ 250 pc) mass concentration, which in turn can dramatically decrease the strength of the bar:
as the central mass increases it can significantly perturb and eventually destroy the orbital structure supporting the bar \citep{hasan90,pfenniger90,friedli93}.
It has been argued, however, that the sole central concentration of mass is not enough to significantly weaken the bar unless its mass is a few percent of the disk mass \citep{shen&sellwood04, athanassoula05}, which is inconsistent with BHs by at least one order of magnitude on the conservative side.
On the other hand, models incorporating the gas response revealed that a frequently overlooked bar-weakening mechanism, namely the transfer of angular momentum between the stellar bar and the infalling gas, can have a significant impact on the bar dissolution, which can take $\lesssim$2 Gyr \citep{bournaud05}.
Interestingly, Bournaud et al. also showed that a noticeable increase in the central mass only happens once the bar has significantly weakened.
These results would imply that a bar-driven build-up of gas in the central regions of the galaxy can be hardly connected to the current strength of the bar.
If said gas was eventually expected to reach and feed the central BH, then it would not surprising that our results show no relation between nuclear activity and the strength of the bar.

In the context of our findings, the above scenario would be particularly appealing.
However, most recent simulations from various groups converge toward long-lived and stable bars.
Models in which bars are destroyed tend to use rigid halos, not allowing for angular momentum redistribution which promotes bar growth \citep{athanassoula02b}.
When live halos are used, neither the central mass concentration nor the transfer of angular momentum from the gas to the stellar bar are able to significantly weaken them \citep{berentzen07, villa-vargas10, kraljic12, athanassoula13}, and therefore bar weakening can be likely ruled out as the cause of the disconnection between bar strength and ongoing nuclear activity.

\subsection{Nuclear Bars and Nuclear Rings}

\citet{shlosman89} proposed a cascade of instabilities in a galaxy as a possible way of fueling BH activity--the ``bars within bars'' scenario.
Gas inflows driven by a large-scale bar would result in a circumnuclear gaseous disk, which could in turn suffer from further instabilities and form a randomly-oriented nested bar within the large-scale bar.
This nuclear bar could drive gaseous material further down into the galactic nucleus and feed an AGN.
{\em HST} observations of nearby Seyfert galaxies, however, have found nuclear bars in only a minority of them \citep{martini01, laine02}.

Nuclear rings can be found in around one fifth of barred galaxies \citep{comeron10}.
They are thought to be signposts of inflowing gas slowing down near the inner Lindblad resonances \citep{simkin80, combes&gerin85, knapen95}.
Their relation to the fueling of nuclear activity could be twofold: as they trace a recent gas inflow to the nuclear regions, nuclear rings could be expected to be more common in active galaxies \citep[e.g.,][]{knapen05}, or alternatively,
they could indicate that the bulk of the inflowing gas is piling up at the resonances, hindering further significant inflows to smaller scales beyond the nuclear ring \citep[e.g.,][]{garcia-burillo05}.
The latest observational results show that the fraction of galaxies with nuclear rings that also exhibit nuclear activity is consistent with the overall fraction of active galaxies in the nearby universe \citep{comeron10}.

We check whether the degree of nuclear activity of those galaxies from our sample with either of these nuclear features differs from the average by resorting to the morphological classifications by \citet{buta10}.
In their study, a preliminary sample of roughly 10\% of the S$^4$G galaxies were classified using the de Vaucouleurs revised Hubble-Sandage system \citep{devaucouleurs59a}, which accounts for the presence of nuclear rings as well as nuclear bars among various other features.
Currently, classifications exist for the bulk of the S$^4$G sample (R. Buta, private communication), and hence we are able to assess whether the presence of any of these nuclear features makes a difference in the nuclear fueling.
We complement the classifications with those in the catalogs on nuclear bars by \citet{laine02} and \citet{erwin04}, and on nuclear rings by \citet{comeron10}. 

In Table \ref{tab1}, we indicate which galaxies present nuclear bars and/or rings based on the classifications mentioned above.
Only six galaxies in our sample have nuclear bars, with their median X-ray luminosities and Eddington ratios being
$L_{\mathrm{X}}$ = $1.5 \times 10^{40}$ erg s$^{-1}$
and
$L_{\mathrm{bol}}$/$L_{\mathrm{Edd}}$ = $10^{-5}$
respectively, meaning higher luminosities than the average, yet their accretion rates are similar to those of the parent sample.
Their median morphological $T$-type of 2.2 indicates a mild preference for earlier-types when compared to the parent sample.
Hence, the detected nuclear bars are preferentially found in more massive, earlier-type spirals, implying more massive BHs which in turn accounts for the higher X-ray luminosities yet ordinary Eddington ratios.

Twelve galaxies in our sample feature nuclear rings, with a median $T$-type of 2.9, typical for galaxies hosting nuclear rings \citep{comeron10}, and consistent with the parent sample.
Their median X-ray luminosities and Eddington ratios are
$L_{\mathrm{X}}$ = $6.5 \times 10^{38}$ erg s$^{-1}$
and
$L_{\mathrm{bol}}$/$L_{\mathrm{Edd}}$ = $10^{-5}$
respectively, suggesting that the level of nuclear activity in galaxies those hosting nuclear rings is not different from those in the general population.

\subsection{How to Sustain Low-luminosity AGN Activity}

Early simulations suggested typical bar-driven gas inflow rates into the inner kiloparsec of the order of 0.1--10 $M_{\odot}$ yr$^{-1}$ \citep{friedli93}.
These numbers were empirically confirmed by \citet{sakamoto99}, who estimated a lower limit for the inflow rate into the central kiloparsec of 0.1--1 $M_{\odot}$ yr$^{-1}$ from their observations of molecular gas on nearby spiral galaxies.
Down to smaller scales, the influence of the non-axisymmetric stellar potential on the gas content of nearby active galaxies has been observed to be efficient at driving the gas down to $\sim$100 parsec at rates of 0.01--50 $M_{\odot}$ yr$^{-1}$ \citep{garcia-burillo05, haan09}.
At these scales, these studies have observed that gas inflows are halted and gravity torques can be positive.
From that point on, other competitive mechanisms such as viscous torques could be responsible of driving gas down to smaller scales and potentially reach the BH \citep[e.g.,][]{combes01}.

The mass accretion rates required to sustain typical low-luminosity AGNs, however, are minuscule in comparison to the previously mentioned bar-driven inflow rates:
LLAGN activity is thought to be the product of BHs being fed through radiatively inefficient accretion flows \citep[for a review, see][]{narayan&mcclintock08}.
In this model of mass accretion, for the typical bolometric luminosities and Eddington ratios of our sample, i.e., $L_{\mathrm{bol}}\sim 10^{40}$ erg s$^{-1}$ and $L_{\mathrm{bol}}/L_{\mathrm{Edd}}\sim 10^{-5}$ respectively, \citet{ho09b} estimates mass accretion rates of the order of $\dot{M} \sim 10^{-6}$--$10^{-5} \,M_{\odot}$ yr$^{-1}$.
In the context of these extremely modest accretion rates, \citet{ho09b} argues that most galaxies have their innermost regions a readily available steady supply of fuel in the form of (1) stellar mass loss from evolved stars, and (2) Bondi accretion of hot gas.
These fuel sources can exceed the estimated BH mass accretion rates by $\sim$2 orders of magnitude, and hence bar-driven gas inflows, while sufficient, might not be a necessary condition to sustain typical low-level AGN activity observed in the nearby universe and could account for the independence between nuclear activity and bar strength found in the present study.

\section{Conclusions}

In this work, we quantified both
the bar strengths of a sample of 41 nearby barred galaxies from {\em Spitzer}/IRAC imaging,
and the level of BH activity using {\em Chandra}/ACIS archival data.
Based on the observational and theoretical evidence that bars drive material toward the central regions of a galaxy,
our goal was to determine whether bar strength has an impact on the level of BH fueling
by investigating possible correlations between different measures of bar strength and AGN activity.
Our findings can be summarized as follows:

\begin{itemize}

\item[1.] We found a nuclear X-ray point source coincident with the NIR position in 31 out of 41 galaxies indicative of ongoing BH fueling.
The median 2--10 keV X-ray luminosity and Eddington ratio of $L_{\mathrm{X}}$ = $4.3 \times 10^{38}$ erg s$^{-1}$ and $L_{\mathrm{bol}}$/$L_{\mathrm{Edd}}$ = $6.9 \times 10^{-6}$ respectively are consistent with the levels of low-luminosity nuclear activity in the nearby universe \citep{ho09b}.
For those sources without detections, upper limits were derived.

\item[2.] We estimated the strength of the stellar bar in two independent ways:
from its structure, as traced by its ellipticity and boxiness, and from its maximum relative gravitational torque.
No significant correlation was found between any of the bar strength proxies and the level of AGN activity:
statistical tests on our sample did not reveal any significant trend between bar strength and BH fueling, irrespective of galaxy luminosity, stellar mass, Hubble type, or bulge size.
This suggests that the strength the stellar bar, and therefore the extent of the bar-driven inflow, is not directly connected with the degree of ongoing BH fueling, at least for the low-luminosity regime probed here.

\item[3.] We checked whether the presence of nuclear rings and/or nuclear bars had any impact on the ongoing BH fueling.
We found that galaxies with nuclear rings show similar levels of nuclear activity compared to the parent sample, while galaxies with nuclear bars tend to have slightly higher luminosities yet ordinary Eddington ratios, mainly because they tend to be found in earlier-type galaxies with higher mass BHs.

\item[4.] Assessing our findings in the broader context of previous results from the literature, we discuss possible scenarios concluding that
(1) because strong bars are not necessarily related to more efficient BH fueling, the mechanisms responsible for LLAGN activity can not be traced on scales larger than a few hundred parsec;
and (2) the mass accretion rates required to sustain LLAGN activity are minuscule in comparison to the observed bar-driven inflow rates, and therefore other sources readily available at the centers of most galaxies must provide a steady supply of fuel without the need of kiloparsec scale inflows.

\end{itemize}

%% Included in this acknowledgments section are examples of the
%% AASTeX hypertext markup commands. Use \url without the optional [HREF]
%% argument when you want to print the url directly in the text. Otherwise,
%% use either \url or \anchor, with the HREF as the first argument and the
%% text to be printed in the second.

\acknowledgments

M.C. thanks Professor Ron Buta for providing morphological classifications for the sample, and the anonymous referee for useful suggestions.
D.A.G. thanks Michael West for useful discussions.
E.A. and A.B. acknowledge the CNES (Centre
National d'Etudes Spatiales - France) for financial support.
We acknowledge financial support to the DAGAL network from the People 
Programme (Marie Curie Actions) of the European Union's Seventh 
Framework Programme FP7/2007-2013/ under REA grant agreement number 
PITN-GA-2011-289313.
This work was co-funded under the Marie Curie Actions of the European
Commission (FP7-COFUND).
%NRAO
The National Radio Astronomy Observatory is a facility of the National Science Foundation operated under cooperative agreement by Associated Universities, Inc.
This research is based in part on observations made with the {\em Spitzer Space Telescope}, and makes use of the NASA/IPAC Extragalactic Database (NED),
both of which are operated by the Jet Propulsion Laboratory, California Institute of Technology under a contract with the National Aeronautics and Space Administration.
We acknowledge the usage of the HyperLeda database (http://leda.univ-lyon1.fr).
This publication makes use of data products from the Two Micron All Sky Survey, which is a joint project of the University of Massachusetts and the Infrared Processing and Analysis Center/California Institute of Technology, funded by the National Aeronautics and Space Administration and the National Science Foundation.

{\it Facilities:} \facility{CXO}, \facility{{\it Spitzer} (IRAC)}.




%\begin{appendix}
\appendix

\begin{figure*}[th]
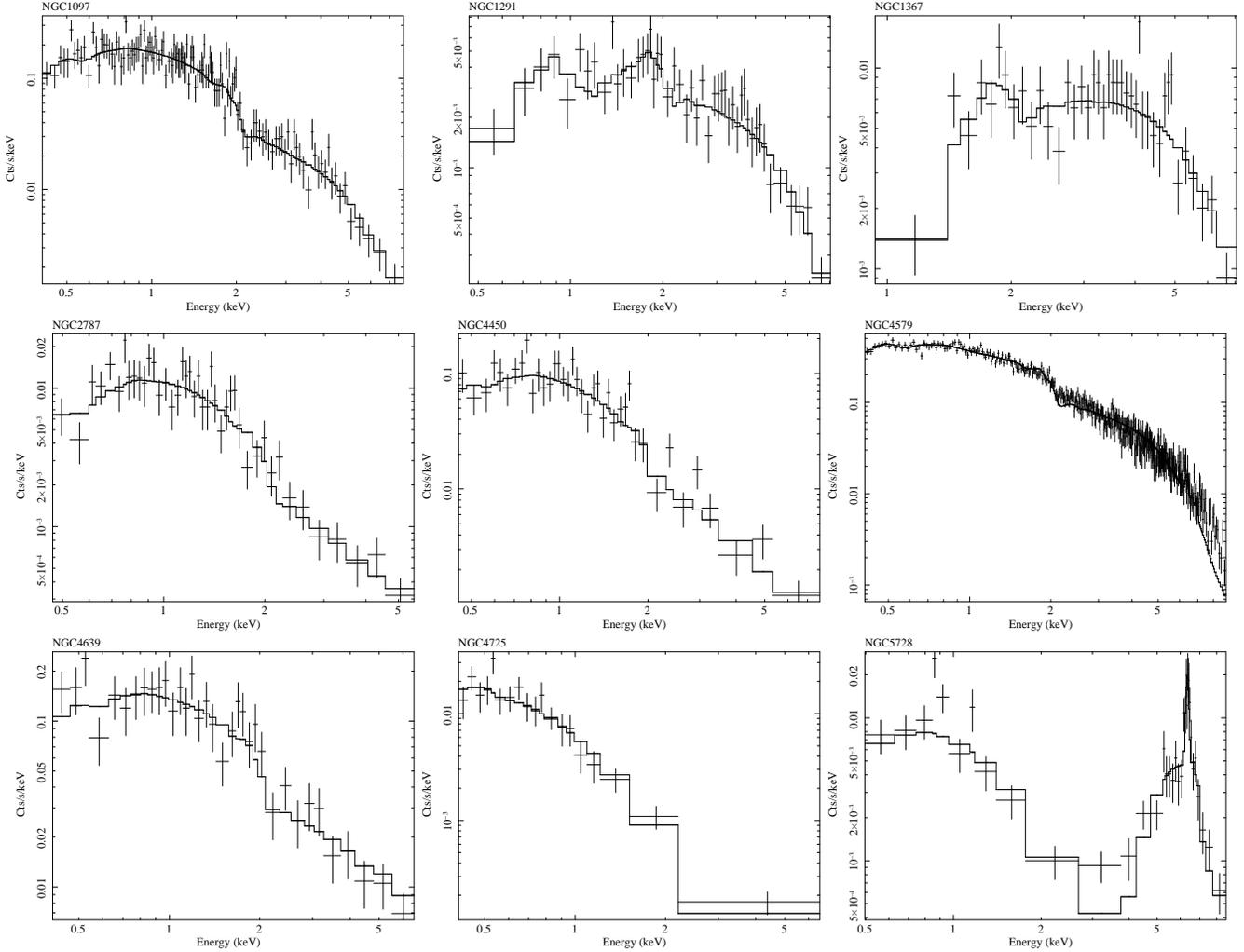

\begin{center}
\includegraphics[width=0.25\textwidth,angle=-90]{NGC1097_po.eps}
\includegraphics[width=0.25\textwidth,angle=-90]{NGC1291_pome.eps}
\includegraphics[width=0.25\textwidth,angle=-90]{NGC1367_po.eps}
\includegraphics[width=0.25\textwidth,angle=-90]{NGC2787_po.eps}
\includegraphics[width=0.25\textwidth,angle=-90]{NGC4450_po.eps}
\includegraphics[width=0.25\textwidth,angle=-90]{NGC4579_2po.eps}
\includegraphics[width=0.25\textwidth,angle=-90]{NGC4639_po.eps}
\includegraphics[width=0.25\textwidth,angle=-90]{NGC4725_po.eps}
\includegraphics[width=0.25\textwidth,angle=-90]{NGC5728_2po.eps}
\caption{Nuclear X-ray spectra and best-fitting models for the subsample of sources with more than 200 counts.\label{fig_spec}}
\end{center}
\end{figure*}

\section{Spectral Modeling}

For the nine sources in our sample with more than 200 net counts, we perform a spectral analysis using {\ttfamily XSPEC} v12.7.0 \citep{xspec}. 
The spectra were binned using the {\ttfamily GRPPHA} task included in {\ttfamily FTOOLS} so that each spectral bin had at least 20 counts, allowing us to use $\chi^2$ statistics to select a best fitting model.
While a single power law model is a good representation of the AGN emission, bright sources with high count rates require a more detailed analysis in order  to properly characterize their nature and disentangle additional components contributing to the observed emission, e.g., thermal plasma emission or a prominent iron line at 6.4 KeV. 
An ensemble of five models, as in \citet{ogm09}, is used to better represent the true nature of the emitting source.
These correspond to:
a power-law model (PL) with intrinsic absorption, accounting for non-thermal AGN emission;
a single-temperature thermal plasma model (MEKAL) to account for emission from unresolved binaries or supernova remnants;
a combined MEKAL+PL model in which the spectrum shows a contribution from both thermal and non-thermal emission mechanisms in the soft and hard X-rays respectively;
a double power-law model (2PL), in which a second power law is used to account for possible AGN continuum emission scattered off the surrounding medium and showing up in the soft X-rays, with both described by the same spectral index;
and a MEKAL+2PL model, similar as the previous one but adding a thermal component also at lower energies.

In order to choose the best model, we require the resulting parameters to have realistic values with a physical meaning, e.g., photon index $\Gamma$ = 0--3 for the PL and temperature $kT$ = 0--2 keV for the MEKAL model.
In the case that multiple models return reasonable parameters, preference is given to the simplest model (ie., the one with the least number of components) by checking that the quality of the fit does not improve significantly by adding additional components using the $F$-test task within {\ttfamily XSPEC}.
In order to discern between models with the same number of components, the one with the $\chi^2_{\nu}$ closest to unity is chosen.
The best-fitting models together with their corresponding parameters are presented in Table \ref{tab5};
the spectra together with the chosen model are shown in Figure \ref{fig_spec};
and our findings are briefly discussed below:

%%%%%%%%%%%%%%%%%%%%%%%
\begin{deluxetable}{llccccc}
\tabletypesize{\scriptsize}
\tablecaption{Best fit models and parameters\label{tab5}}
\tablehead{Galaxy & Model & $N_{\mathrm{H,1}}$ & $N_{\mathrm{H,2}}$ & $\Gamma$ & $kT$ & $\chi^{2}_{\nu}$\\
 & & ($10^{22}$ cm$^{-2}$) & ($10^{22}$ cm$^{-2}$) & & (keV) & \\
 (1) & (2) & (3) & (4) & (5) & (6)
}
\startdata
NGC 1097&PL& $  0.04^{+  0.04}_{-  0.03}$ & ... & $  1.64^{+  0.13}_{-  0.12}$ & ... &  0.96 \\
NGC 1291&MEPL& $  0.67^{+  0.27}_{-  0.35}$ & $  1.87^{+  1.02}_{-  0.63}$ & $  1.89^{+  0.63}_{-  0.39}$ &  $0.18^{+  0.10}_{-  0.06}$ &  0.80 \\
NGC 1367&PL& $  2.42^{+  1.02}_{-  0.46}$ & ... & $  1.15^{+  0.55}_{-  0.13}$ & ... &  0.97 \\
NGC 2787&PL& $  0.10^{+  0.08}_{-  0.08}$ & ... & $  2.29^{+  0.40}_{-  0.37}$ & ... &  1.13 \\
NGC 4450&PL& $  0.04^{+  0.07}_{-  0.04}$ & ... & $  2.18^{+  0.37}_{-  0.26}$ & ... &  1.24 \\
NGC 4579&2PL& $  1.89^{+  0.31}_{-  0.32}$ & $  0.01^{+  0.02}_{-  0.01}$ & $  1.61^{+  0.06}_{-  0.06}$ & ... &  1.40 \\
NGC 4639&PL& $  0.03^{+  0.07}_{-  0.03}$ & ... & $  1.34^{+  0.29}_{-  0.24}$ & ... &  0.88 \\
NGC 4725&PL& $  0.01^{+  0.08}_{-  0.01}$ & ... & $  3.34^{+  1.22}_{-  0.28}$ & ... &  0.98 \\
NGC 5728&2PL& $  0.01^{+  0.81}_{-  0.01}$ & $100.62^{+ 17.45}_{- 22.03}$ & $  2.41^{+  0.40}_{-  0.37}$ & ... &  1.56

\enddata
\tablecomments{Column(1): galaxy name;
Column (2): best-fitting model, in these cases either an absorbed power-law (PL) or a double power-law (2PL) model;
Columns (3) and (4) : HI column densities of model components;
Column (5): spectral photon index;
Column (6): temperature of the thermal plasma;
Column (7): reduced $\chi^2$.
}
\end{deluxetable}
%%%%%%%%%%%%%%%%%%%%%%%

{\it NGC 1367} -
We report spectral modeling of the X-ray nuclear source in NGC 1367 for the first time, finding that its nuclear spectrum is best-fit by a single power-law with a rather hard photon index.
Regarded as a non-active galaxy, it was observed with {\em Chandra} to study SN2005ke \citep{immler06}.

{\it NGC 1097, NGC 2787, and NGC 4450} -
The X-ray nuclei of these galaxies are best-fit by single power-law models characteristic of LLAGNs.
Optically, the nuclei of these three galaxies have been found to belong to the LINER family by \citet{phillips84}, \citet{heckman80}, and \citet{ho97a} respectively, an indication of the likely non-thermal nature of the nuclear emission on this class of active nuclei.

{\it NGC 1291} -
The nucleus of this galaxy is a LINER as well \citep{smith07}, and is best-fit by a MEKAL+PL model, in which the thermal plasma component accounts for the soft X-ray excess.
Model parameters are in agreement with the detailed study of the X-ray source population of this galaxy by \citet{luo12}.

{\it NGC 4579} -
Similarly, NGC 4579 also features a LINER nucleus \citep{stauffer82} which is best-fit by a 2PL model.
\citet{eracleous02} modeled the nuclear X-ray source as a single unabsorbed power law, while \citet{ogm09} find that the nuclear source is best-fit by a MEKAL+PL model. The main differences seem to arise in the soft X-ray part of the spectrum, and the 2--10 keV luminosities derived from these studies and ours agree among each other.

{\it NGC 4639 and NGC 4725} -
Both of these galaxies host Seyfert nuclei \citep{ho97a} and are best-fit by single power-law models, in agreement with their known AGN nature. 

{\it NGC 5728} -
The X-ray nuclear source in NGC 5728 is best-fit by the 2PL model together with a Gaussian to account for the strong FeK$\alpha$ feature at 6.4 KeV.
The hard power law shows an absorption two orders of magnitude larger when compared to the rest of the sources from our sample, and just at the limit for being considered Compton-thick, at $N_{\mathrm{H,2}} \sim 10^{24}$ cm$^{-2}$.
This is in agreement with the value already reported by \citet{comastri10} from {\em Suzaku} observations. \\

%\end{appendix}

%\bibliographystyle{apj}
%\bibliography{mauricio}
\twocolumngrid

\end{document}